# Enhancing Fault Detection and Isolation in All-Electric Auxiliary Power Unit (APU) Gas Generator by Utilizing Starter/Generator Signal

**Haotian Mao [1,*], Khashayar Khorasani [2] and Yingqing Guo [1,*]**

[1] School of Power and Energy, Northwestern Polytechnical University, Xi'an 710072, China
[2] Department of Electrical and Computer Engineering, Concordia University, Montreal, QC H3G 1M8, Canada; kash@ece.concordia.ca
* Correspondence: alexmao@mail.nwpu.edu.cn (H.M.); yqguo@nwpu.edu.cn (Y.G.)

**Abstract**

This study proposes a novel paradigm for enhancing the fault detection and isolation (FDI) of gas generators in all-electric auxiliary power unit (APU) by utilizing shaft power information from the starter/generator. First, we conduct an investigation into the challenges and opportunities for FDI that are brought about by APU electrification. Our analysis reveals that the electrification of APUs opens new possibilities for utilizing shaft power estimates from starters/generators to improve gas generator FDI. We then provide comprehensive theoretical and analytical evidence demonstrating why, how, and to what extent the shaft power information from the starter/generator can fundamentally enhance the estimation accuracy of system states and health parameters of the gas generator, while also identifying key factors influencing these improvements in FDI performance. The effectiveness of the proposed paradigm and its theoretical foundations are validated through extensive Monte Carlo simulation runs. The research findings provide a unique perspective in addressing three fundamental questions—why joint fault diagnosis of the starter/generator and gas generator in all-electric APUs is essential, how it can be implemented, and what factors determine its effectiveness—thereby opening up promising new avenues for FDI technologies in all-electric APU systems.

## 1. Introduction

To address the escalating need for air travel and its imminent economic and environmental ramifications, the most viable solution at present is aircraft electrification. While the notion of an all-electric aircraft (AEA) remains unrealized, the concept of more-electric aircraft (MEA) has elicited considerable interest. In these systems, electrical counterparts supplement or entirely supplant the cumbersome, inefficient hydraulic and pneumatic systems that are characteristic of conventional aircraft [1–3]. Due to this trend, the functionality and structure of the auxiliary power unit (APU) have undergone significant changes, transitioning from the conventional APU that provided aerodynamic, hydraulic, and electrical power to an all-electric APU that solely provides electrical power [4,5].

The all-electric APU has been implemented on the Boeing 787 aircraft. The key difference between the traditional APU and the all-electric APU is that the latter is equipped with a high-power electric starter/generator and eliminates the pneumatic and hydraulic functions and their associated components of the APU [4].

The estimation and diagnosis of APU failures are critically important for flight safety, as the APU is closely related to the key energy supply of the aircraft and serves as a vital component for starting the main engines. Additionally, estimating degradation and health conditions are beneficial for achieving condition-based maintenance (CBM), which can reduce maintenance cost. Consequently, numerous studies have been conducted on APU failure and health condition estimation, including those focused on gas generators and starters/generators [6–13].

The authors of [8] proposed a method for identifying the health status of gas generator components based on the recognition of system performance parameters. This method trains a classification algorithm for each system component to identify its health status. As compared to the training strategy based on a single fault assumption, the training strategy considering multiple component faults provides more accurate diagnostic estimation. The authors of [9] proposed a method that combines long short-term memory (LSTM) network and support vector regression (SVR) techniques with a Kalman filter to estimate key performance parameters of gas generators. The authors of [10] utilized the random forest method to establish a performance baseline model for an APU and then obtained estimates of health indices characterizing the performance degradation of an in-service APU based on the performance baseline model. The authors of [14] proposed a multi-time-window convolutional bidirectional long short-term memory neural network for fault detection and isolation (FDI) of APUs in civil aircraft.

The authors of [11] presented a time-to-failure analysis of a more-electric aircraft starter/generator relying on the physics of failure approach, with a special focus on turn-to-turn short-circuit faults. The authors of [12] utilized an extended Kalman filter (EKF) to estimate inter-turn short-circuit faults in the APU aircraft starter/generator. The authors of [13] proposed a method based on stacked autoencoders (SAEs) and support vector data description (SVDD) for detecting rectifier faults in the aircraft starter/generator.

However, the above studies have been conducted focusing on individual components, such as the gas generator or starter/generator. This approach of conducting fault estimation research on each component separately is feasible for traditional APUs. Nevertheless, for all-electric APUs, this isolated approach is not the optimal solution, due to strong coupling relationship between the starter/generator and gas generator of an all-electric APU. On one hand, this coupling causes a malfunction of either component to severely affect the operation of the other, posing a challenge to our fault estimation results. On the other hand, one can utilize this coupling relationship to introduce new external information into the fault estimation of individual components, thereby enhancing the fault estimation capability, which presents opportunities for improving the fault estimation results. This paper will conduct an in-depth investigation into this opportunity, exploring the enhancement of our capability in estimating gas generator faults through the utilization of starter/generator signals.

The organization of this paper is as follows. Section 2 introduces our research objectives and the all-electric APU and conducts an in-depth analysis of the opportunities and challenges in fault diagnosis that are brought by the electrification of APUs. Section 3 provides the premises and assumptions for our proposed methodology. Section 4 presents the posterior estimates using and not using the shaft power estimate from the starter/generator. Section 5 explores the differences in estimation accuracy between the two estimates, aiming to demonstrate how the information of shaft power from the starter/generator can improve the accuracy of the gas generator fault estimation and identifies factors that affect the magnitude of this improvement. Section 6 presents the Monte Carlo simulation results, validating the improvement in estimation accuracy that is achieved through incorporating the shaft power data from the starter/generator. Additionally, it demonstrates that employing a more accurate model or estimating additional health parameters can significantly enhance the utility of shaft power information in fault estimation, as well as in FDI objectives.

## 2. Statement of Problem and Research Objectives

In this section, we will first introduce an all-electric APU and then present our analysis of the opportunities and challenges that the electrification of the APU brings to the fault estimation problem. Finally, we will briefly outline the research objectives of our study.

Our research focuses on an all-electric APU, primarily comprising a gas generator and a high-power starter/generator. The gas generator operates as a turboshaft, producing minimal thrust while primarily generating power via its output shaft [15]. The starter/generator, on the other hand, functions as a brushless wound-field synchronous generator (BWFSG) [16]. The operational mechanism of an all-electric APU involves the gas generator transmitting power to the starter/generator via the drive shaft and gearbox, thereby enabling the starter/generator to generate electricity and supply power to the aircraft [17].

The all-electric APU is an evolution from the traditional APU. Conventionally, aircraft requires the APU to supply electrical, pneumatic, and hydraulic energy. Therefore, the shaft power output of the traditional APU is required to drive the starter/generator for electrical energy, the compressor for pneumatic energy, and the equipment for hydraulic energy. However, with the electrification of aircraft, more-electric aircraft no longer require the APU to provide pneumatic and hydraulic energy but only require electrical energy from the APU. This implies that the all-electric APU, as opposed to the traditional APU, only needs the shaft power to drive a high-power starter/generator.

Taking the Hamilton Sundstrand APS5000 all-electric APU used in the Boeing 787 as an example, the structure of a typical all-electric APU is shown in Figure 1 [18]. Using the Honeywell 131-9 series APU employed in Boeing 737 and Airbus A320 aircraft as an example, the structure of a typical traditional APU is shown in Figure 2 [19]. The gas generators in both systems are single-shaft turboshaft engine structures. The principles and structures of the traditional APU and the all-electric APU are compared in the form shown in Figure 3, where $P$ represents power and $N_{speed}$ denotes rotor speed. The dashed line represents the components that are reduced in an all-electric APU as compared to the traditional APU. It can be seen that the all-electric APU eliminates two components that convert rotational power into pneumatic and hydraulic energy, resulting in a simpler structure. This simpler coupling and more direct relationship between the gas generator and starter/generator brings about both opportunities and challenges for fault estimation and FDI in the APU.

On one hand, this establishes a strong coupling relationship between the starter/generator and the gas generator, causing their faults to have significant impacts on each other and rendering starter/generator faults more critical [8,20]. This presents a new challenge for our FDI approach. On the other hand, it also offers opportunities to leverage this relationship to enhance our FDI methodology and performance. By utilizing information from one system, one can improve the estimates of the other system. This paper will focus on this opportunity and specifically illustrate how information from the starter/generator can be used to enhance the FDI performance of the gas generator.

Given that the energy output of a gas generator in an all-electric APU is solely reserved for the electrical part of the generator, an opportunity arises for us to precisely estimate the power output (shaft power) of the gas generator, thereby improving the accuracy of the fault diagnosis scheme in the gas generator.

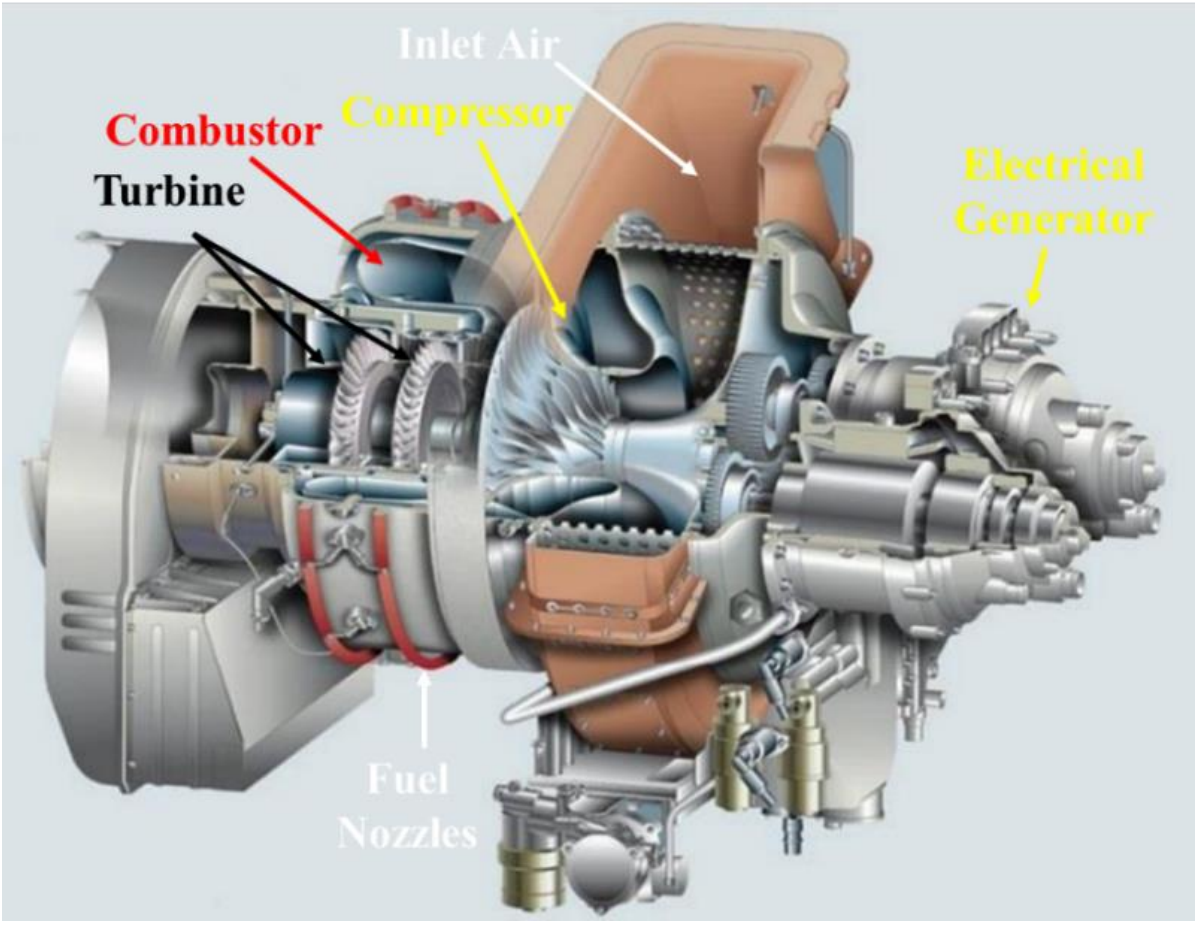

**Figure 1.** Structural diagram of APS5000 all-electric APU [18].

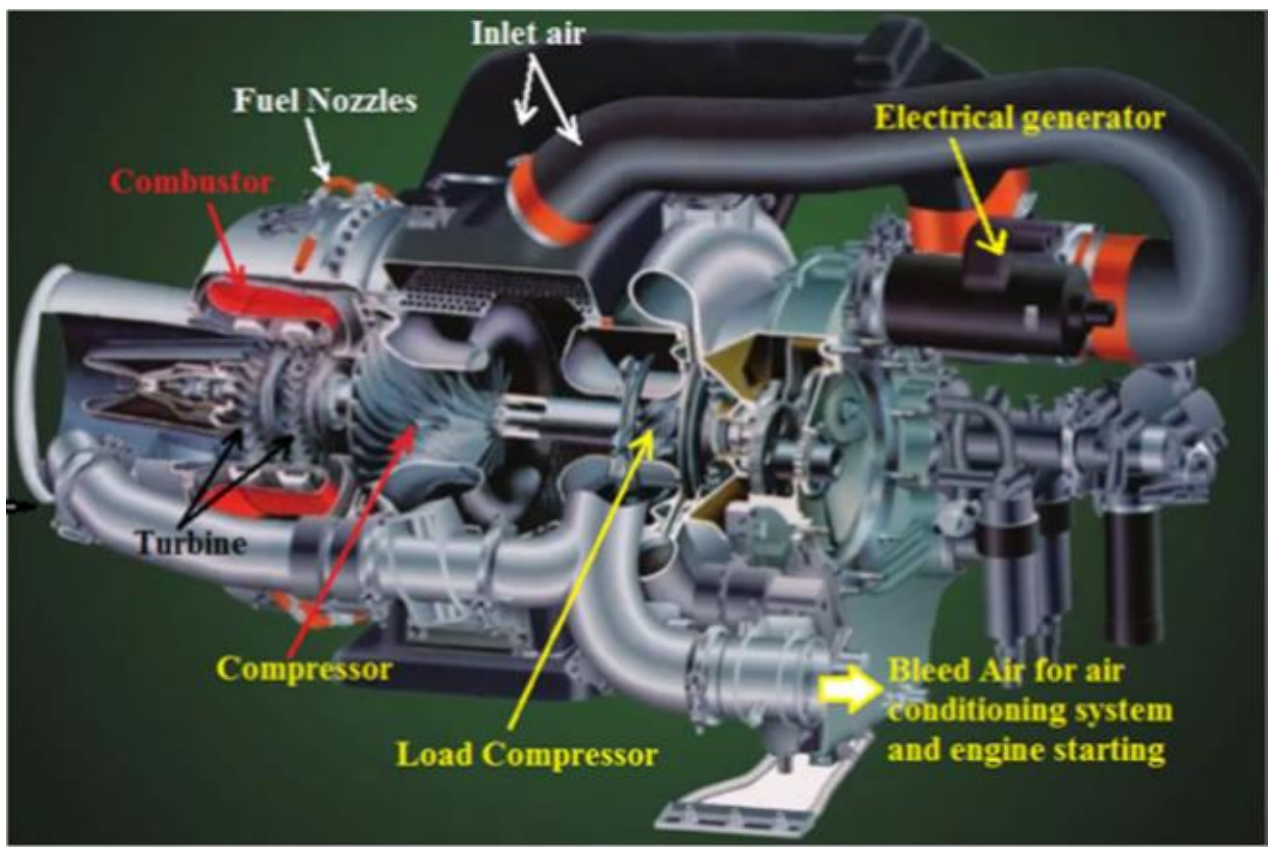

**Figure 2.** Structural diagram of 131-9B traditional APU [19].

The energy output of a traditional APU gas generator equals the sum of the energy consumed by the starter/generator that supplies electrical energy to the aircraft, the compressor that supplies the pneumatic energy, and the hydraulic pump that supplies the hydraulic energy. As a component of the electrical system, the power consumed by the starter/generator can be estimated through measurable electrical measurements. However, the energy consumed by the compressor and hydraulic pump is influenced by numerous factors, and obtaining accurate relevant parameters is a highly challenging task. Therefore, making a precise estimate of their power consumption is quite challenging. Moreover, due to the compact structure of the APU, installing torque or power sensors inside it poses significant challenges, which has always made estimating the power of the gas generator a complex task [21,22].

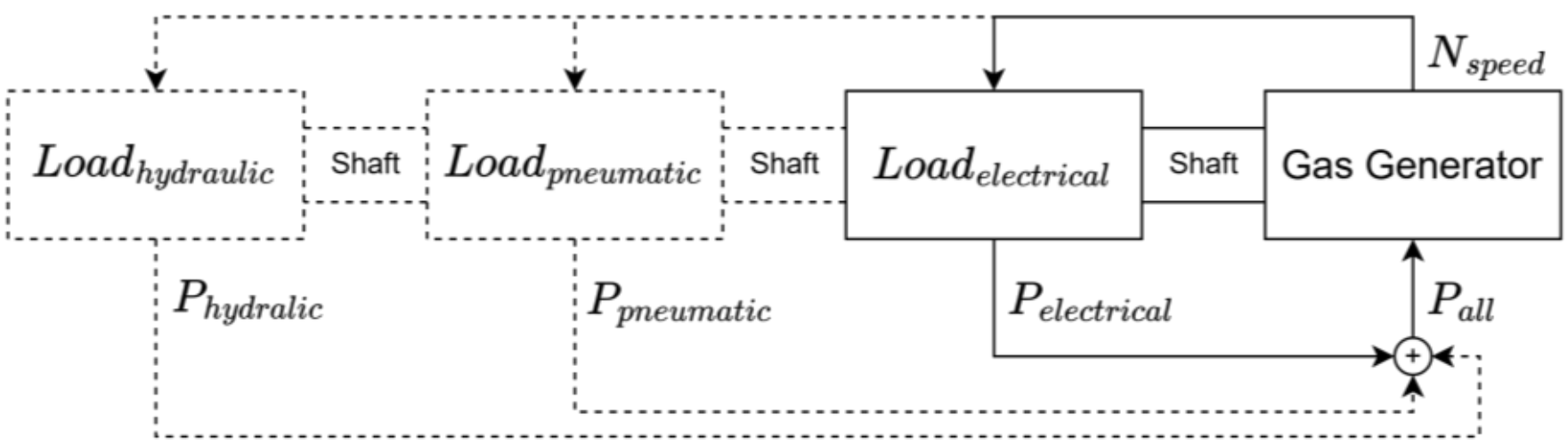

**Figure 3.** Comparison between the traditional APU and an all-electric APU.

The advent of an all-electric APU offers a solution to the above problem. Without the compressor and hydraulic pump, one can estimate the power of the starter/generator with relative precision, thereby obtaining a fairly accurate estimate of power output of the gas generator. This information on the shaft power of gas generator can significantly enhance the estimation accuracy of states and the health parameters of the gas generator.

The feasibility of the above concept has already been demonstrated in other similar electro-mechanical systems, such as the hybrid electric vehicle and wind power generation industries. A range-extended electric vehicle (REEV) is an electric car powered by a battery and equipped with an on-board range extender generator, also known as an auxiliary power unit (REEV-APU). When the battery charge level is low, the range extender automatically starts and charges the battery. The REEV-APU consists of an engine and a generator, where the engine is solely responsible for driving the generator, which in turn supplies electrical power to the vehicle. Researchers in this field have conducted extensive studies on how to utilize the generator to estimate the engine's shaft power and utilize power estimation to improve the parameter estimation and performance of the REEV-APU [23–25]. Similarly, the wind power generation system primarily comprises a generator and a mechanical wind turbine, where the wind turbine, driven by wind power, serves as the prime mover that drives the generator to produce electricity. Researchers in this field have also conducted a substantial amount of research on how to utilize the generator's signals, including shaft power information, to estimate the parameters, monitor the conditions, and diagnose faults in the wind turbine [26,27].

In this study, our aim and objective are to investigate why, how, and to what extent shaft power estimates from the starter/generator can enhance system state and health parameter estimation in an all-electric APU gas generator, thereby improving FDI performance.

## 3. Preliminaries

In this section, we introduce the gas generator model and describe its characteristics. We will then introduce the methodological framework that is adopted in our research.

### *3.1. Gas Generator Model*

We assume that the dynamics of the gas generator can be modeled by a turboshaft aero-engine linear system, which is typically expressed in the following state-space representation [28–32]:

$$\begin{aligned} x(t_{k+1}) &= Ax(t_k) + Bu(t_k) + E\theta(t_k) + FPe(t_k) + w(t_k) \\ y(t_k) &= Cx(t_k) + Du(t_k) + G\theta(t_k) + v(t_k) \end{aligned} \tag{1}$$

where $u$ represents the gas generator input, y represents the measurements; $\theta$ denotes the health parameters that represent gas generator faults or degradation; $Pe$ represents the shaft power; $x$ represents the state variables; $A$, $B$, $C$, $D$, $E$, $F$, and $G$ represent the known matrices of the state-space model corresponding to the system; $n$ and $m$ represent the dimensions of each matrix (e.g., $n_A \times m_A$ for matrix $A$), and $w$ and $v$ represent two independent Gaussian white noise processes with zero mean, corresponding to process and measurement noise, respectively.

In the context of a gas turbine system, it is a reasonable assumption that the power output from the shaft affects all rotor speeds and state parameters. Each measurement is influenced by at least one state variable, and each health parameter impacts at least one measurement. To formalize these assumptions, it can be noted that all elements in the column vector $F$ are non-zero. Moreover, the matrix $C$ does not contain any rows composed entirely of zero elements, and the matrix $G$ does not have any columns composed entirely of zero elements.

To illustrate the rationality of the above three assumptions, consider gas-path faults in gas turbine engines as an example. The first assumption is reasonable because shaft power extraction affects the energy available for rotor acceleration, thereby influencing rotor speed and subsequently other engine state variables [33,34]. The second assumption is justified since rotor speed, as a fundamental operating parameter, drives the thermodynamic processes that determine temperature and pressure measurements throughout the gas-path [35–37]. The third assumption is supported by the physical nature of gas-path health parameters, which represent component performance degradations (such as compressor fouling or turbine blade erosion) that alter flow conditions and thereby affect measurable quantities like pressure, temperature, and speed [30,32,38,39]. These physical relationships are consistently reflected in linear engine models reported in the literature [32,35,36,40].

Furthermore, given the operational characteristics of an all-electric APU, it is hypothesized that the health parameters demonstrate gradual evolutionary patterns, with their specific degradation trajectories being inherently uncertain a priori. Simultaneously, the shaft power, predominantly determined by electrical loading conditions, is subject to substantial stochastic variations and may undergo rapid transient changes [41,42]. Within this framework, it is assumed that the starter/generator system provides high-fidelity shaft power estimates and associated covariance matrices at each temporal increment, with these estimates conforming to a Gaussian probability distribution. This assumption allows us to abstract away from the technical complexities of obtaining precise shaft power estimates, such as noise mitigation strategies and signal loss compensation mechanisms, thereby maintaining the investigative focus on the fundamental relationship between the shaft power estimation and the gas generator FDI capabilities.

**Remark 1.** *Equation (1) represents a generalized gas generator system model that incorporates both health parameters and shaft power [30–32]. This linear system representation, combined with typical all-electric APU characteristic assumptions, enables us to analyze how shaft power information fundamentally impacts fault estimation and diagnosis, rather than limiting the analysis to specific APU structures. Moreover, the adoption of linear system modeling is also justified by the gas generator's inherently slow and relatively linear dynamic response characteristics, which has been widely validated in relevant literature [43–46].*

It is important to note that in this paper, we do not delve into the specific technical details of how to accurately obtain shaft power information, including considerations such as shaft torsion effects, gearbox power losses, and the conversion from electrical to mechanical power. Instead, we focus on establishing the theoretical foundation by assuming that accurate shaft power information is available, analyzing how this information fundamentally affects gas generator fault estimation and diagnosis capabilities, and investigating what factors determine its effectiveness. The practical aspects of obtaining precise shaft power measurements, including dealing with various engineering challenges and constraints, will be systematically addressed in our subsequent research.

### *3.2. Proposed Methodology*

The premise of a linear system allows the application of linear state estimation techniques to estimate system states, health parameters, and to assess the influence of shaft power estimates, which are obtained from the starter/generator.

This study introduces two estimates: posterior estimates with no shaft power information (PENSs) and posterior estimates with shaft power information (PESs). We will demonstrate that the PENSs represent the posterior estimates of health parameters and system states in the absence of accurate shaft power estimates at each step. Conversely, the PESs represent the posterior estimates of the variables when accurate shaft power estimates, derived from the starter/generator, are available at each step.

Comparing the accuracy of the PESs and PENSs allows one to assess the impact of shaft power estimates from starter/generator on these estimates. By analyzing this impact, one can determine how the shaft power estimates derived from the starter/generator can enhance the FDI in gas generators and identify factors that affect the magnitude of this enhancement. Finally, we employ Monte Carlo simulations to validate our proposed methodology and to analyze how the significance of shaft power information changes in response to variations in the accuracy of the model that is used for estimation and the number of health parameters.

## 4. Two Posterior Estimates: With and Without Shaft Power Information from Starter/Generator Signals

In this section, we introduce two different estimation scenarios. The first scenario involves estimating system states and health parameters with the aid of an accurate shaft power estimate from the starter/generator. The second scenario proceeds without such information. For these two scenarios, we then define two corresponding posterior estimates: posterior estimates with shaft power information (PESs) and posterior estimate with no shaft power information (PENSs). Subsequently, we propose two algorithms specifically designed to compute and derive these two estimates.

### *4.1. Definition of Two Scenarios and Two Posterior Estimates*

Considering the system that is described by Equation (1), we assume a given Gaussian distribution for the initial system states $x(t_0)$ and health parameters $\theta(t_0)$. The inputs $u(t_k)$ and measurements $y(t_k)$ are also available at each time step. The posterior estimates of the health parameters at the time step $t_{k-1}$, denoted by $\hat{\theta}(t_{k-1})$, are utilized to compute the prior estimate $\tilde{\theta}(t_k)$, at the subsequent time step $t_k$, which is expressed as $\tilde{\theta}(t_k) = \hat{\theta}(t_{k-1}) + w^h(t_{k-1})$, where $w^h(t_{k-1})$ represents noise that accounts for the uncertainty in health parameters due to the underlying dynamically slow degradation processes.

*Scenario 1 (PES)*: A given Gaussian estimate $Pe(t_k)$ and the corresponding covariance $P^{Pe}(t_k)$ for the shaft power from the starter/generator is available at each time step. The PES is defined as the posterior estimate of system states and health parameters at each time step given the conditions that are outlined above.

*Scenario 2 (PENS)*: The pre-set values $PeT$ and $P^{PeT}$ are utilized as the estimated value and the covariance of the shaft power for each time step, respectively, where $PeT$ represents the shaft power value under a typical operating condition and $P^{PeT}$ is a sufficiently large quantity. The PENS is defined as the posterior estimate of system states and health parameters at each time step given the conditions outlined above.

The health parameters, as assumed above, change slowly. Thus, the value at the next time step is almost the same as that of the previous one. Consequently, we choose to use the posterior estimate from the last time step and noise to derive the prior estimate for the next time step in both scenarios [28,29]. Due to absence of shaft power data in Scenario 2, we selected $PeT$ and $P^{PeT}$ as estimation values and covariances of the shaft power to address the uncertainty associated with the shaft power, aiming for accurate posterior estimates.

**Remark 2.** *The above two estimation scenarios, for the first time, clearly define the estimation process and the achievable posterior estimates under two conditions, that is with/without shaft power information. This novel framework enables the evaluation of shaft power information's influence through comparative analysis of the estimation accuracy and the FDI performance between the PESs and PENSs.*

*4.2. Algorithms for the Two Posterior Estimates*

According to the characteristics of the gas generator and the definition of Scenario 1, one can incorporate health parameters into the system states, as outlined in Equation (2), to obtain the PES [28,29], that is

$$\begin{aligned} x^{aug}(t_{k+1}) &= A^{aug}x^{aug}(t_k) + B^{aug}u(t_k) + F^{aug}Pe(t_k) + w^{aug}(t_k) \\ y(t_k) &= C^{aug}x^{aug}(t_k) + Du(t_k) + v(t_k) \end{aligned} \tag{2}$$

where

$$A^{aug} = \begin{bmatrix} A & E \\ 0 & I \end{bmatrix},\ B^{aug} = \begin{bmatrix} B \\ 0 \end{bmatrix},\ C^{aug} = \begin{bmatrix} C & G \end{bmatrix},\ F^{aug} = \begin{bmatrix} F \\ 0 \end{bmatrix},$$

$$Q^{aug} = E\left[w^{aug}(t_k)w^{aug}(t_k)^T\right],\ R = E\left[v(t_k)v(t_k)^T\right],$$

$$w^{aug}(t_k) = \begin{bmatrix} w(t_k) & 0 \\ 0 & w^h(t_k) \end{bmatrix},\ x^{aug}(t_k) = \begin{bmatrix} x(t_k) \\ \theta(t_k) \end{bmatrix},$$

and the superscript $aug$ denotes the augmented variable.

Following the above augmentation, state estimation scheme can be utilized to achieve the PES. By definition of Scenario 1 and Equation (2), the prior estimates and covariance update process for achieving the PES can be described by

$$\tilde{x}^{aug}(t_k) = A^{aug}\hat{x}^{aug}(t_{k-1}) + B^{aug}u(t_{k-1}) + F^{aug}Pe(t_{k-1}) \tag{3}$$

$$\tilde{P}^{aug}(t_k) = A^{aug}\hat{P}^{aug}(t_{k-1})A^{augT} + F^{aug}P^{Pe}(t_{k-1})\left(F^{aug}\right)^T + Q^{\text{aug}} \tag{4}$$

where the superscript ~ represents parameters related to the prior estimate and $\tilde{P}^{aug}$ represents the covariance of the prior estimate.

When we have the above prior estimates and covariance, one can employ Kalman filter theory to obtain the PES. The corresponding Kalman gain update equation, posterior estimate, and the covariance update equations for achieving the PES can be expressed by [47]

$$K^{aug}(t_k) = \tilde{P}^{aug}(t_k)\left(C^{aug}\right)^T\left(C^{aug}\tilde{P}^{aug}(t_k)\left(C^{aug}\right)^T + R\right)^{-1} \tag{5}$$

$$\hat{x}^{aug}(t_k) = \tilde{x}^{aug}(t_k) + K^{aug}(t_k)\left(y(t_k) - C^{aug}\tilde{x}^{aug}(t_k) - Du(t_k)\right) \tag{6}$$

$$\hat{P}^{aug}(t_k) = \left(I - K^{aug}(t_k)C^{aug}\right)\tilde{P}^{aug}(t_k)\left(I - K^{aug}(t_k)C^{aug}\right)^T + K^{aug}(t_k)RK^{aug}(t_k)^T \tag{7}$$

where the superscript ^ represents the parameters related to the posterior estimate, $\hat{P}^{aug}$ denotes the posterior estimate covariance matrix, and $K^{aug}$ represents the Kalman gain.

By combining the above prior and posterior update Equations (3)–(7), Algorithm 1 can be derived for obtaining the PES as shown in Algorithm 1.

To obtain the PENS, it is also practical to incorporate health parameters and shaft power into the system states, as demonstrated in Equation (2).

By definition, in Scenario 2, the prior estimates and the covariance update process for achieving the PENS can be described by

$$\tilde{x}^{aug}(t_k) = A^{aug}\hat{x}^{aug}(t_{k-1}) + B^{aug}u^{aug}(t_{k-1}) + F^{aug}PeT \tag{8}$$

$$\tilde{P}^{aug}(t_k) = A^{aug}\hat{P}^{aug}(t_{k-1})\left(A^{aug}\right)^T + F^{aug}P^{PeT}\left(F^{aug}\right)^T + Q^{aug} \tag{9}$$

The corresponding Kalman gain update equation, posterior estimate, and the covariance update equation for achieving the PENS can also be expressed by Equations (5)–(7) above.

**Algorithm 1:** Posterior estimates with shaft power information (PES).

| **Algorithm 1:** Posterior estimates with shaft power information (PES) | | |
|---|---|---|
| **Input** | : | Engine augmented state space matrix: $A^{aug}$, $B^{aug}$, $C^{aug}$, $D^{aug}$, $F^{aug}$; Posterior state estimate and covariance at $t_{k-1}$: $\hat{x}^{aug}(t_{k-1})$, $\hat{P}^{aug}(t_{k-1})$; Shaft power estimate from starter/generator and corresponding covariance: $Pe(t_{k-1})$, $P^{Pe}(t_{k-1})$; |
| **Output** | : | Posterior state estimate and covariance at $t_k$: $\hat{x}^{aug}(t_k)$, $\hat{P}^{aug}(t_k)$; |
| **Step 1** | : | **Predict** 1. Predict the prior estimate $\tilde{x}^{aug}(t_k)$ based on Equation (3); 2. Predict the prior estimate covariance $\tilde{P}^{aug}(t_k)$ based on Equation (4); |
| **Step 2** | : | **Update** 1. Obtain the Kalman gain $K^{aug}(t_k)$ based on Equation (5); 2. Update posterior state estimate $\hat{x}^{aug}(t_k)$ based on Equation (6); 3. Update posterior state estimate covariance $\hat{P}^{aug}(t_k)$ based on Equation (7); |

By combining the prior Equations (8) and (9) and posterior Equations (5)–(7), Algorithm 2 can be derived to obtain the PENS as follows (Algorithm 2):

**Algorithm 2:** Posterior estimates with no shaft power information (PENS).

| **Algorithm 2:** Posterior estimates with no shaft power information (PENS) | | |
|---|---|---|
| **Input** | : | Engine augmented state space matrix: $A^{aug}$, $B^{aug}$, $C^{aug}$, $D^{aug}$, $F^{aug}$; Posterior state estimate and covariance at $t_{k-1}$: $\hat{x}^{aug}(t_{k-1})$, $\hat{P}^{aug}(t_{k-1})$; |
| **Output** | : | Posterior state estimate and covariance at $t_k$: $\hat{x}^{aug}(t_k)$, $\hat{P}^{aug}(t_k)$; |
| **Step 1** | : | **Predict** 1. Predict the prior estimate $\tilde{x}^{aug}(t_k)$ based on Equation (8); 2. Predict the prior estimate covariance $\tilde{P}^{aug}(t_k)$ based on Equation (9); |
| **Step 2** | : | **Update** 1. Obtain the Kalman gain $K^{aug}(t_k)$ based on Equation (5); 2. Update posterior state estimate $\hat{x}^{aug}(t_k)$ based on Equation (6); 3. Update posterior state estimate covariance $\hat{P}^{aug}(t_k)$ based on Equation (7); |

In this section, we define the estimation scenarios based on whether the shaft power information from the starter/generator is available or not. Subsequently, we present detailed algorithms for obtaining the posterior estimates and their corresponding covariance under these two scenarios, namely, the PES and PENS.

## 5. Impact of Shaft Power Information from the Starter/Generator on the Gas Generator Fault Estimation Accuracy

This section evaluates the impact of shaft power information from the starter/generator on the estimation accuracy through a comparison between the PES and PENS. To facilitate this comparison, we develop MPESs (modified posterior estimates with shaft power information), a modified version of the PES model, and demonstrate its equivalence to the PENS model. This equivalence enables us to use MPESs as a bridge for effectively comparing PESs and PENSs. The results demonstrate that incorporating shaft power information significantly improves the estimation precision of each gas generator state and health parameter. This improvement becomes more pronounced under two conditions, namely, when using a more accurate state transition model or when estimating a larger number of health parameters.

### *5.1. Modified Posterior Estimate with Shaft Power Information (MPES) and Its Equivalence with PENS*

After substituting Equation (9) for Equation (4) in the PES, we derive a modified estimate that is referred to as the MPES, and its algorithm is outlined as the Algorithm 3.

**Algorithm 3:** Modified posterior estimates with shaft power information (MPES).

| **Algorithm 3:** Modified posterior estimates with shaft power information (MPES) | | |
|---|---|---|
| **Input** | : | Engine augmented state space matrix:<br>$A^{aug}$, $B^{aug}$, $C^{aug}$, $D^{aug}$, $F^{aug}$;<br>Posterior state estimate and covariance at $t_{k-1}$:<br>$\hat{x}^{aug}(t_{k-1})$, $\hat{P}^{aug}(t_{k-1})$;<br>Shaft power estimate from starter/generator:<br>$Pe(t_{k-1})$; |
| **Output** | : | Posterior state estimate and covariance at $t_k$:<br>$\hat{x}^{aug}(t_k)$, $\hat{P}^{aug}(t_k)$; |
| **Step 1** | : | **Predict**<br>1. Predict the prior estimate $\tilde{x}^{aug}(t_k)$ based on Equation (3);<br>2. Predict the prior estimate covariance $\tilde{P}^{aug}(t_k)$ based on Equation (9); |
| **Step 2** | : | **Update**<br>1. Obtain the Kalman gain $K^{aug}(t_k)$ based on Equation (5);<br>2. Update posterior state estimate $\hat{x}^{aug}(t_k)$ based on Equation (6);<br>3. Update posterior state estimate covariance $\hat{P}^{aug}(t_k)$ based on Equation (7); |

The substitution that is outlined in Algorithm 3 reveals that the MPES is derived by leveraging the precise estimate from the starter/generator, without utilizing the corresponding covariance. Instead, the pre-set large value $P^{PeT}$, used in Algorithm 2 for computing the PENS, is employed as the covariance. Consequently, the MPES represents estimates that are obtained by utilizing the accurate shaft power estimates from the starter/generator, albeit with minimal belief. Following this, we now introduce Theorem 1 to elucidate the equivalence between the MPES and PENS schemes.

**Theorem 1.** *The estimation covariance and the Kalman gain in Algorithm 3 for achieving the MPES is equal to that in Algorithm 2 for achieving the PENS at each time step following* $t_0$*, as denoted by* $\hat{P}^{aug,MPES}(t_k)=\hat{P}^{aug,PENS}(t_k)$ *and* $K^{aug,MPES}(t_k)=K^{aug,PENS}(t_k)$*, where the superscripts* $MPES$ *and* $PENS$ *correspond to the estimates and covariance of the MPES and PENS, respectively.*

**Proof.** Since both algorithms for the PENS and MPES use Equations (4), (5), and (7), respectively, to compute the prior covariance $\tilde{P}^{aug,PENS}(t_k)$ and $\tilde{P}^{aug,MPES}(t_k)$, the Kalman gain $K^{aug,PENS}(t_k)$ and $K^{aug,MPES}(t_k)$, and the posterior covariance $\hat{P}^{aug,PENS}(t_k)$ and $\hat{P}^{aug,MPES}(t_k)$, it follows that $\tilde{P}^{aug,PENS}(t_k)=\tilde{P}^{aug,MPES}(t_k)$, $K^{aug,PENS}(t_k)=K^{aug,MPES}(t_k)$, and $\hat{P}^{aug,PENS}(t_k)=\hat{P}^{aug,MPES}(t_k)$ when $\hat{P}^{aug,PENS}(t_{k-1})=\hat{P}^{aug,MPES}(t_{k-1})$, provided that the system parameters and health parameters, as well as the measurements and inputs, are the same.

Thus, it can be inferred through recursive reasoning that $\hat{P}^{aug,MPES}(t_k)=\hat{P}^{aug,PENS}(t_k)$ and $K^{aug,MPES}(t_k)=K^{aug,PENS}(t_k)$ at each time step following $t_0$ when the system parameters, the initial Gaussian distribution of the system states and health parameters, as well as the measurements and inputs are the same.

This completes the proof of the theorem. □

Here, we have demonstrated the equality of the covariance and the Kalman gain between the PENS and MPES. Next, we will demonstrate that the estimates of the PENS and MPES are also the same under the conditions.

**Theorem 2.** *Assuming that* $P^{aug,PENS}_{[i,i]}(t_k)$ *is bounded for any* $i$ *when* $P^{PeT}\to+\infty$*, where* $i$ *represents any row of the matrix and the subscript* $[:,:]$ *refers to the matrix element at the corresponding row and column, then it follows that as* $P^{PeT}\to+\infty$*, the difference between* $\hat{x}^{aug,PENS}(t_k)$ *and* $\hat{x}^{aug,MPES}(t_k)$ *converges to zero following the initial time* $t_0$.

**Proof.** By expanding Equations (4) and (7), one can express the posterior estimate variance of the PENS as follows:

$$\begin{aligned}\hat{P}^{aug,PENS}(t_k)=&\left(I-K^{aug,PENS}(t_k)C^{aug}\right)F^{aug}P^{PeT}\left(F^{aug}\right)^T\\&\left(I-K^{aug,PENS}(t_k)C^{aug}\right)^T+\left(I-K^{aug,PENS}(t_k)C^{aug}\right)\\&\left(A^{aug}\hat{P}^{aug,PENS}(t_{k-1})\left(A^{aug}\right)^T+Q^{\text{aug}}\right)\\&\left(I-K^{aug,PENS}(t_k)C^{aug}\right)^T+K^{aug,PENS}(t_k)RK^{aug,PENS}(t_k)^T\end{aligned} \tag{10}$$

It follows that all the three parts in Equation (10) are positive semi-definite matrices. Considering that when $\hat{P}^{aug,PENS}_{[i,i]}$ is bounded for any $i$, it follows that $\left(I-K^{aug,PENS}(t_k)C^{aug}\right)F^{aug}\to0$ as $P^{PeT}\to+\infty$.

Using Equation (6) and Theorem 1 above, and considering the same inputs and measurements, one can derive

$$\begin{aligned}\hat{x}^{aug,PENS}(t_k)-\hat{x}^{aug,MPES}(t_k)=&\left(I-K^{aug,PENS}(t_k)C^{aug}\right)A^{aug}\\&\left(\hat{x}^{aug,PENS}(t_{k-1})-\hat{x}^{aug,MPES}(t_{k-1})\right)\\&+\left(I-K^{aug,PENS}(t_k)C^{aug}\right)F^{aug}\left(PeT-Pe(t_{k-1})\right)\end{aligned} \tag{11}$$

Based on the result that $\left(I - K^{aug,PENS}(t_k)C^{aug}\right)F^{aug} \to 0$ as $P^{PeT} \to +\infty$, it can be inferred that if $\hat{x}^{aug,PENS}(t_{k-1}) - \hat{x}^{aug,MPES}(t_{k-1}) \to 0$, thus $\hat{x}^{aug,PENS}(t_k) - \hat{x}^{aug,MPES}(t_k) \to 0$ as $P^{PeT} \to +\infty$.

Then, through recursive reasoning, one can derive that as $P^{PeT} \to +\infty$, $\hat{x}^{aug,PENS}(t_k) - \hat{x}^{aug,MPES}(t_k) \to 0$ for each time step following the initial time $t_0$, under the assumptions of the theorem. This completes the proof of the theorem. □

Theorem 2 implies the practical equivalence of estimates between the PENS and MPES. This is due to the fact that the $P^{PeT}$ is a large number, as previously described, and when PENS is effective (signifying that the estimation error is not significantly large, i.e., $P^{aug,PENS}_{[i,i]}(t_k)$ is not a large value for any $i$), then $\hat{x}^{aug,PENS}(t_k)$ and $\hat{x}^{aug,MPES}(t_k)$ are almost the same.

By combining the covariance equivalence that is presented in Theorem 2 with the equivalence of estimates that is outlined in Theorem 3, one can conclusively establish the equivalence between the PENS and the MPES.

In this subsection, we proposed the MPES and demonstrated the equivalence of the PENS and MPES. Subsequently, we will assess and compare the estimation accuracy of the PENS and PES by examining the accuracy between the MPES and PES.

### *5.2. Comparative Estimation Accuracy Analysis of PENS and PES*

In this section, we will demonstrate and derive the differences in accuracy between the PES and PENS. Given the demonstrated equivalence between the PENS and MPES in the previous subsection, we can assess the estimation accuracy of the PES relative to the PENS through a comparison with the MPES.

**Theorem 3.** *Assuming that the shaft power estimates and corresponding covariances from starter/generator, which are provided to the PES, are consistent with the actual system and* $P^{PeT}(t_k) > P^{Pe}(t_k)$ *in each time step, then it follows that* $\varepsilon_i^{aug,MPES}(t_k) \geq \varepsilon_i^{aug,PES}(t_k)$ *for any* $i$ *and time step following* $t_0$*, where* $\varepsilon_i^{aug,MPES}(t_k) = E\left[\left(\hat{x}_i^{aug,MPES}(t_k) - x_i^{aug}(t_k)\right)^2\right]$ *and* $\varepsilon_i^{aug,PES}(t_k) = E\left[\left(\hat{x}_i^{aug,PES}(t_k) - x_i^{aug}(t_k)\right)^2\right]$*, representing the expected squared estimation errors for the* $i$ *th estimate of the MPES and PES, respectively.*

**Proof.** We first denote $P_\varepsilon^{aug,MPES}(t_k)$ and $P_\varepsilon^{aug,PES}(t_k)$ as the covariance of estimation error (CEE) [47] of the MPES and PES as follows:

$$P_\varepsilon^{aug,MPES}(t_k) = E\left[\left(\hat{x}^{aug,MPES}(t_k) - x^{aug}(t_k)\right)\left(\hat{x}^{aug,MPES}(t_k) - x^{aug}(t_k)\right)^T\right] \tag{12}$$

$$P_\varepsilon^{aug,PES}(t_k) = E\left[\left(\hat{x}^{aug,PES}(t_k) - x^{aug}(t_k)\right)\left(\hat{x}^{aug,PES} - x^{aug}(t_k)\right)^T\right] \tag{13}$$

where subscript $\varepsilon$ denotes the expected squared estimation error matrix.

The proof can now be transformed into demonstrating that $P_{\varepsilon,[i,i]}^{aug,MPES}(t_k) \geq P_{\varepsilon,[i,i]}^{aug,PES}(t_k)$ for any given $i$ and time step subsequent to $t_0$, under the stipulated assumptions.

Given that the parameters that are provided to compute the PES are all consistent with the actual system, it follows that the posterior covariance of the PES is equal to its CEE, namely $P_\varepsilon^{aug,PES}(t_k) = \hat{P}^{aug,PES}(t_k)$ holds for any time step following $t_0$. Consequently, based on Equation (7), the CEE of the PES, as denoted by $P_\varepsilon^{aug,PES}(t_k)$, can be expressed as follows [47]:

$$\begin{aligned} P_{\varepsilon}^{aug,PES}(t_k) &= \hat{P}^{aug,PES}(t_k) \\ &= \left(I - K^{aug,PES}(t_k)C^{aug}\right)\tilde{P}^{\varepsilon^{aug,PES}}(t_k)\left(I - K^{aug,PES}(t_k)C^{aug}\right)^T \\ &\quad + K^{aug,PES}(t_k)RK^{aug,PES}(t_k)^T \\ &= \left(I - K^{aug,PES}(t_k)C^{aug}\right)\tilde{P}^{aug,PES}(t_k)\left(I - K^{aug,PES}(t_k)C^{aug}\right)^T \\ &\quad + K^{aug,PES}(t_k)RK^{aug,PES}(t_k)^T \end{aligned} \tag{14}$$

where

$$\tilde{P}_{\varepsilon}^{aug,PES}(t_k) = A^{aug}P^{\varepsilon^{aug,PES}}(t_{k-1})\left(A^{aug}\right)^T + F^{aug}P^{Pe}(t_{k-1})\left(F^{aug}\right)^T + Q^{\text{aug}}$$

Due to the lack of actual shaft power distribution information in the calculations for the MPES, the posterior estimate of the MPES is not the CEE of the MPES. The CEE of the MPES is presented as follows [48]:

$$\begin{aligned} P_{\varepsilon}^{aug,MPES}(t_k) &= \left(I - K^{aug,MPES}(t_k)C^{aug}\right)\tilde{P}_{\varepsilon}^{aug,MPES}(t_k)\left(I - K^{aug,MPES}(t_k)C^{aug}\right)^T \\ &\quad + K^{aug,MPES}(t_k)RK^{aug,MPES}(t_k)^T \end{aligned} \tag{15}$$

where

$$\tilde{P}_{\varepsilon}^{aug,MPES}(t_k) = A^{aug}\hat{P}_{\varepsilon}^{aug,MPES}(t_{k-1})\left(A^{aug}\right)^T + F^{aug}P^{Pe}(t_{k-1})\left(F^{aug}\right)^T + Q^{\text{aug}}$$

By utilizing Equations (14) and (15), the expression for $P_{\varepsilon}^{aug,MPES}(t_k) - P_{\varepsilon}^{aug,PES}(t_k)$ can be derived as shown in Equation (16), as follows:

$$\begin{aligned} P_{\varepsilon}^{aug,MPES}(t_k) - P_{\varepsilon}^{aug,PES}(t_k) &= \left(K^{aug,MPES}(t_k) - K^{aug,PES}(t_k)\right)\left(C^{aug}\tilde{P}^{aug,PES}(t_k)\left(C^{aug}\right)^T + R\right) \\ &\quad \left(K^{aug,MPES}(t_k) - K^{aug,PES}(t_k)\right)^T \\ &\quad + \left(I - K^{aug,MPES}(t_k)C^{aug}\right)\left(\tilde{P}_{\varepsilon}^{aug,MPES}(t_k) - \tilde{P}_{\varepsilon}^{aug,PES}(t_k)\right) \\ &\quad \left(I - K^{aug,MPES}(t_k)C^{aug}\right)^T \end{aligned} \tag{16}$$

From Equation (16), it can be observed that when $\tilde{P}_{\varepsilon}^{aug,MPES}(t_k) \geq \tilde{P}_{\varepsilon}^{aug,PES}(t_k)$, the right-hand side of Equation (16) consists of two positive semi-definite matrices, resulting in $P_{\varepsilon}^{aug,PES}(t_k) \geq P_{\varepsilon}^{aug,MPES}(t_k)$, where $A \geq B$ represents that $A - B$ is a positive semi-definite matrix, because we have

$$\begin{aligned} &\tilde{P}_{\varepsilon}^{aug,MPES}(t_k) - \tilde{P}_{\varepsilon}^{aug,PES}(t_k) \\ &= A^{aug}\left(P_{\varepsilon}^{aug,MPES}(t_{k-1}) - P_{\varepsilon}^{aug,PES}(t_{k-1})\right)\left(A^{aug}\right)^T \\ &\quad + F^{aug}\left(P^{PeT} - P^{Pe}(t_{k-1})\right)\left(F^{aug}\right)^T \end{aligned} \tag{17}$$

Therefore, through the application of recursive reasoning, it can be concluded that, given $P^{PeT} \geq P^{Pe}(t_k)$ and the same actual initial distribution implying $P_{\varepsilon}^{aug,MPES}(t_0) = P_{\varepsilon}^{aug,PES}(t_0)$, along with other assumptions outlined in the theorem, the inequality $\tilde{P}_{\varepsilon}^{aug,MPES}(t_k) \geq \tilde{P}_{\varepsilon}^{aug,PES}(t_k)$ is maintained at any time step following $t_0$. Hence, based on Equation (16), one can derive that $P_{\varepsilon}^{aug,MPES}(t_k) \geq P_{\varepsilon}^{aug,PES}(t_k)$ holds under the same conditions.

Based on the properties of positive semi-definite matrices and the definition of the CEE, one can conclude that $\varepsilon_i^{aug,MPES}(t_k) \geq \varepsilon_i^{aug,PES}(t_k)$ for any index $i$ and for all time steps subsequent to $t_0$, given the assumptions that are specified in the theorem. This completes the proof of the theorem. □

We next explore the conditions under which equality is consistently achieved in the inequality $\varepsilon_i^{aug,MPES}(t_k) \geq \varepsilon_i^{aug,PES}(t_k)$. From Equation (16), it follows that for the inequality $\varepsilon_i^{aug,MPES}(t_k) \geq \varepsilon_i^{aug,PES}(t_k)$ to consistently hold, the $i$ th row of the matrix $K^{aug,MPES}(t_k) - K^{aug,PES}(t_k)$ must be zero. Referring to Equation (5), it is observed that if the matrices $F$ and $C$ do not contain any rows entirely composed of zero elements, and matrix $G$ does not have any columns entirely composed of zero elements, then the matrix $K^{aug,MPES}(t_k) - K^{aug,PES}(t_k)$ will not have any rows consisting solely of zero elements. Therefore, by combining the equivalence between the PENS and MPES as demonstrated in Theorems 1 and 2, we can conclude that $\varepsilon_i^{aug,PENS}(t_k) > \varepsilon_i^{aug,PES}(t_k)$ for any $i$ under the given estimation scenarios.

Building on the conclusion and the definitions provided in Section 4 for Scenarios 1 and 2, as well as for the PES and PENS, we demonstrate that the use of shaft power information from the starter/generator can improve the accuracy of estimating each system state and health parameter. This improvement will consequently enhance the achievable FDI capabilities.

From the proofs of Theorems 1–3, it can be observed that the improvement in the estimation accuracy for each state and health parameter, due to the shaft power information, is attributable to the intrinsic characteristics of the all-electric APU gas generator. The influence of the shaft power on each state enhances the estimation of each system state. The fact that each measurement is influenced by at least one state, and that, in turn, each health parameter can affect at least one measurement, improves the estimation of each health parameter.

Additionally, the proofs of Theorems 1–3 reveal that the capability of shaft power estimates from the starter/generator to enhance estimation accuracy lies in providing more accurate prior estimates. This improvement becomes more pronounced with the enhancement of the accuracy of the model state transition representation. On the other hand, as the number of health parameters increases, the uncertainty and difficulty of the estimation also increases, thereby elevating the importance of accurate prior estimates. This amplifies the advantages of estimates that utilize shaft power information.

Consequently, as the accuracy of the state transition representation or the number of health parameters in the gas generator model increases, the significance of the improvement in the estimation accuracy of system states and health parameters, enabled by the shaft power information from the starter/generator, becomes more pronounced. A detailed demonstration of this is provided in Section 6.

In this section, we have demonstrated that incorporating shaft power estimates from the starter/generator can improve the estimation accuracy of each state and health parameter of the gas generator, thereby enhancing the achievable FDI performance. We also analyzed the relationship between this improvement and the unique characteristics of the all-electric APU. Additionally, we examined the aspects that can influence the degree of achievable performance improvement. The subsequent section presents simulation results to further substantiate and demonstrate our proposed methodology.

## 6. Simulation Results

In this section, we utilize Monte Carlo simulations to demonstrate the impact of shaft power estimates from the starter/generator on the FDI performance of an all-electric APU gas generator. Furthermore, we investigate the correlation between this impact, model accuracy, and the number of health parameters.

First, we introduce the model of the gas generator that is used in our study. Subsequently, we present and compare the confusion matrices for the FDI in scenarios both with and without shaft power estimates derived from the starter/generator. These comparisons support our conclusion that the incorporation of shaft power information significantly enhances the achievable FDI performance of gas generators.

We also present comparison results for the FDI after adjusting the noise level in the gas generator model and the number of health parameters to be estimated. The results indicate that improvements in the FDI performance are more pronounced by utilizing a more accurate model or a greater number of health parameters.

In this study, gas-path faults are selected as the research focus, and a corresponding gas generator fault model is developed as follows:

$$\begin{aligned} & x^{gt}(t_{k+1}) - x_{ss}^{gt}(t_k) \\ & = A^{gt}\left(x^{gt}(t_k) - x_{ss}^{gt}(t_k)\right) + B^{gt}\left(u^{gt}(t_k) - u_{ss}^{gt}(t_k)\right) \\ & + E^{gt}\left(\theta^{gt}(t_k) - \theta_{ss}^{gt}(t_k)\right) + F^{gt}\left(Pe^{gt}(t_k) - Pe_{ss}^{gt}(t_k)\right) + w^{gt}(t_k) \end{aligned} \tag{18}$$

$$\begin{aligned} & y^{gt}(t_k) - y_{ss}^{gt}(t_k) \\ & = C^{gt}\left(x^{gt}(t_k) - x_{ss}^{gt}(t_k)\right) + D^{gt}\left(u^{gt}(t_k) - u_{ss}^{gt}(t_k)\right) \\ & + G^{gt}\left(\theta^{gt}(t_k) - \theta_{ss}^{gt}(t_k)\right) + v^{gt}(t_k) \end{aligned} \tag{19}$$

where

$$x^{gt} = N,\ \theta^{gt} = \left[e^c, f^c, e^t, f^t\right]^T,\ y^{gt} = [N_m, T_{3m}, P_{3m}, T_{5m}]^T,\ A^{gt} = 0.9962,$$

$$B^{gt} = 4.735 \times 10^3,$$

$$C^{gt} = \begin{bmatrix} 1.000 & 0.01057 & 0.03051 & -0.005151 \end{bmatrix}^T,$$

$$D^{gt} = [0 \quad 2.553\times10^3 \quad 7.786\times10^3 \quad 8.444\times10^3]^T,$$

$$E^{gt} = [187.7 \quad 370.4 \quad -57.70 \quad -24.16],\ F^{gt} = -0.2898,$$

$$G^{gt} = \begin{bmatrix} 0 & 0 & 0 & 0 \\ -368.3 & 238.7 & 174.0 & -176.7 \\ -173.2 & 164.0 & 1.025\times10^3 & -1.299\times10^3 \\ -238.3 & -443.7 & -341.0 & 38.79 \end{bmatrix},$$

and the subscript $ss$ represents the reference steady-state value, $N$ represents the rotor speed (single shaft), $N_m$ represents the measured rotor speed, $T_{3m}$ represents the measured compressor outlet temperature, $P_{3m}$ represents the measured compressor outlet pressure, $T_{5m}$ represents the measured turbine outlet temperature, and $e^c$, $f^c$, $e^t$, and $f^t$, respectively, represent the degradation of compressor efficiency, compressor flow capability, turbine efficiency, and turbine flow capability.

The above linear model formulation is a typical representation of single-shaft all-electric APU gas generator gas-path fault models commonly employed in the industry [40,49,50]. This model was derived by linearizing a component-level nonlinear all-electric APU model developed using GasTurb [51]. The component-level model's parameters and design are based on the Hamilton Sundstrand APS 5000, which is a mainstream single-shaft all-electric APU widely used in the current field [18]. The modeling approach employs GasTurb-based dynamic link library technology [52–54], which enables the simulation of gas-path faults in the gas generator. The model's validity has been confirmed through validation using widely recognized industry software and relevant literature. To maintain the shaft speed at its nominal value, a proportional–integral (PI) controller was also designed and incorporated.

Gas-path faults represent a critical area of focus in the field, as such faults significantly affect the efficiency and flow capacity of major rotating components [40,49,50]. The linearized model described above constitutes a typical all-electric APU gas generator gas-path fault model, with rotational speed as the state variable and the health parameters $e^c$, $f^c$, $e^t$, and $f^t$ used to quantify the severity of gas-path faults. These four parameters represent the efficiency and flow factors of the compressor and turbine, respectively, as defined in Equations (20) and (21). In a healthy state, these parameters have a baseline value of 1. Deviations from this baseline can indicate changes in efficiency and flow due to gas-path degradation. The gas-path degradation of compressors and turbines is primarily caused by issues such as blade corrosion, erosion, fouling, and tip wear, among others, which lead to alterations in blade parameters and overall performance. Typically, compressor degradation results in decreased efficiency and reduced flow, whereas turbine degradation is characterized by decreased efficiency albeit with increased flow [55–59]. Specifically, we have

$$e^c = \eta^c / \eta^{c-n}, \quad e^t = \eta^t / \eta^{t-n} \tag{20}$$

$$f^c = m^c / m^{c-n}, \quad f^t = m^t / m^{t-n} \tag{21}$$

where $\eta^c$ and $\eta^t$ denote the actual efficiencies of the compressor and turbine, respectively, $\eta^{c-n}$ and $\eta^{t-n}$ represent their nominal efficiencies, $m^c$ and $m^t$ indicate the actual flow rates, and $m^{c-n}$ and $m^{t-n}$ correspond to the nominal flow rates.

It is evident that this model conforms to the model structure and satisfies all assumptions outlined in Section 3, thus enabling its application for validating the effectiveness of our analytical conclusions, while also confirming the rationality of our related assumptions. Based on the values of these health parameters, degradation is categorized into four distinct health conditions, namely healthy, minor fault, medium fault, and severe fault. These conditions are detailed in Table 1.

It is worth noting that the above classification of healthy, minor fault, medium fault, and severe fault states does not carry any special physical considerations. This classification serves to differentiate fault severity levels, enabling us to evaluate PES-based and PENS-based approaches through their ability to distinguish different fault magnitudes. Combined with the assessment of fault estimation accuracy, this provides a comprehensive evaluation of both fault estimation and the FDI capabilities of PES-based and PENS-based approaches. This helps validate our theoretical framework by examining whether fault estimation and diagnosis with shaft power information indeed outperforms those without it, and whether the factors affecting this shaft power information-induced improvement align with our analysis. Such fault classification and testing methodology has been widely adopted in various aero gas turbine fault estimation and diagnosis studies to compare the estimation and diagnostic capabilities of different approaches [60–63].

**Table 1.** Health condition classification.

| | Healthy | Minor Fault | Medium Fault | Severe Fault |
|---|---|---|---|---|
| Compressor efficiency factor $e^c$ | 0.98–1 | 0.96–0.98 | 0.94–0.96 | 0.92–0.94 |
| Compressor flow factor $e^t$ | 0.98–1 | 0.96–0.98 | 0.94–0.96 | 0.92–0.94 |
| Turbine efficiency factor $f^c$ | 0.98–1 | 0.96–0.98 | 0.94–0.96 | 0.92–0.94 |
| Turbine flow factor $f^t$ | 1–1.02 | 1.02–1.04 | 1.04–1.06 | 1.06–1.08 |
| Sample Size | 100 | 100 | 100 | 100 |

***Case 1*:** We set measurement noise $v^{gt}$ and process noise $w^{gt}$ to be 0.5% of the reference steady-state values of states and measurements, respectively. We conducted 100 Monte Carlo simulation runs for each condition, as shown in Table 1. Each simulation began completely healthy, namely all health parameters were set to 1, and gradually degraded to the target condition. We used the PES and PENS algorithms as described in Section 4 to estimate health parameters. Subsequently, we performed an FDI scheme based on these estimation results to identify the health condition associated with each degradation. Figure 4 illustrates the dynamic estimation results from a representative simulation sample. Table 2 shows the estimation accuracy of the PES and PENS, where RMSE represents the root mean square error. Figures 5–8 present FDI confusion matrices based on the PES and PENS for $e^c$, $f^c$, $e^t$, and $f^t$, respectively.

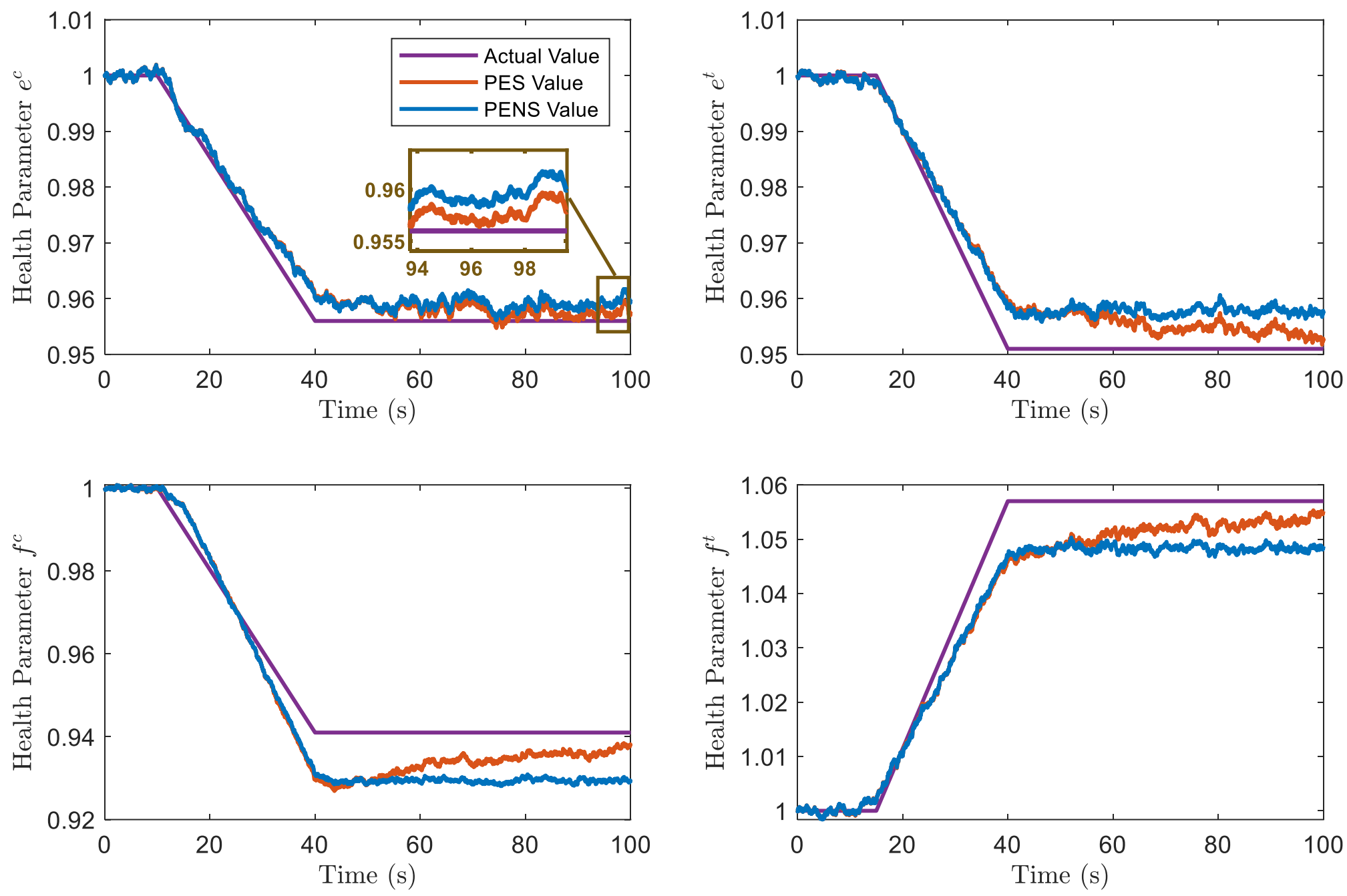

**Figure 4.** Case 1 dynamic estimation results for PES and PENS.

Figure 4 and Table 2 demonstrate that the PES significantly outperforms the PENS in terms of estimation accuracy for each state and health parameter. Also, from Figures 5–8, we can observe that FDI utilizing the PES demonstrates superior performance as compared to FDI based on the PENS, particularly as fault severity escalates.

**Table 2.** Comparison of accuracy between the PES and PENS under Case 1.

| | RMSE of PES | RMSE of PENS | Accuracy Improvement of the PES over PENS |
|---|---|---|---|
| $N_L$ | 105.8 | 121.8 | 13.13% |
| $e^c$ | $2.417\times10^{-3}$ | $3.207\times10^{-3}$ | 24.63% |
| $f^c$ | $8.749\times10^{-3}$ | $1.194\times10^{-2}$ | 26.74% |
| $e^t$ | $5.042\times10^{-3}$ | $6.835\times10^{-3}$ | 26.23% |
| $f^t$ | $6.604\times10^{-3}$ | $9.006\times10^{-3}$ | 26.67% |

For the healthy condition, the FDI based on both the PES and PENS demonstrates comparable performance. This can be attributed to the fact that a completely healthy system does not exhibit any dynamic changes in health parameters due to degradation. Each parameter consistently maintains a value of 1, thereby reducing the complexity of the estimation and resulting in similar FDI performance.

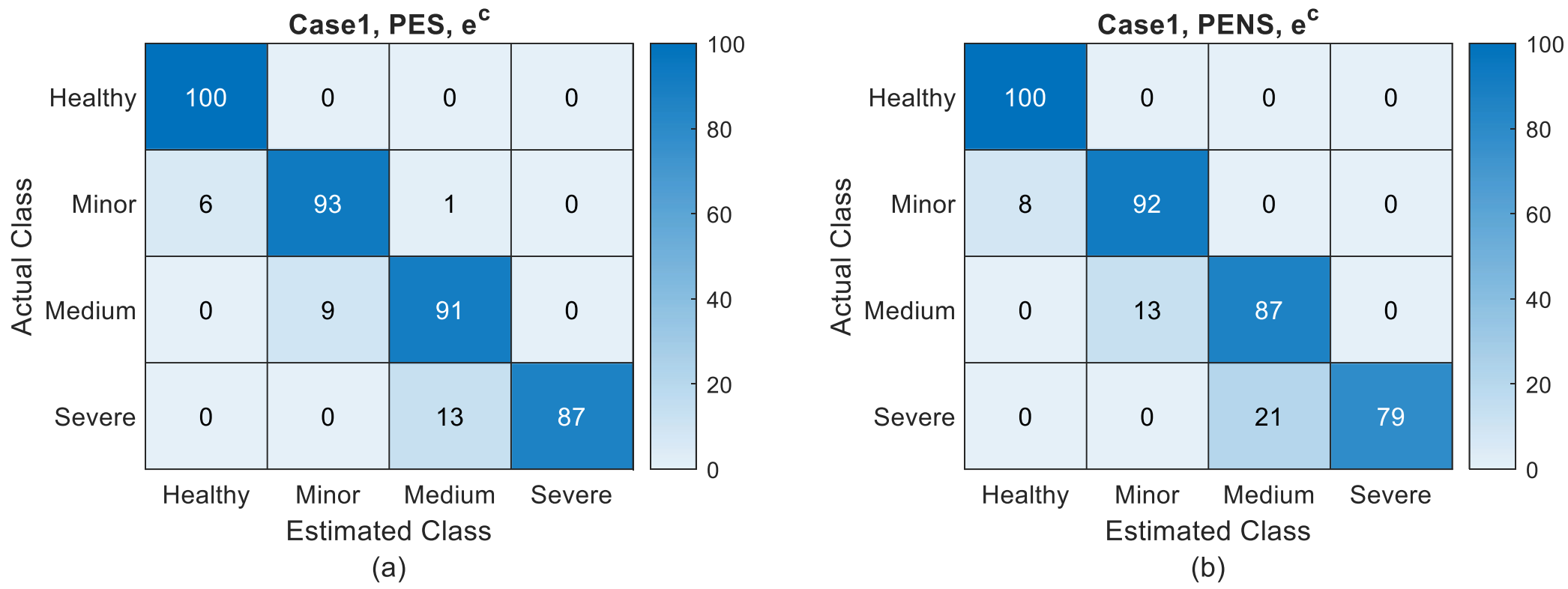

**Figure 5.** Confusion matrix for $e^c$ based on the PES (**a**)/PENS (**b**) under Case 1.

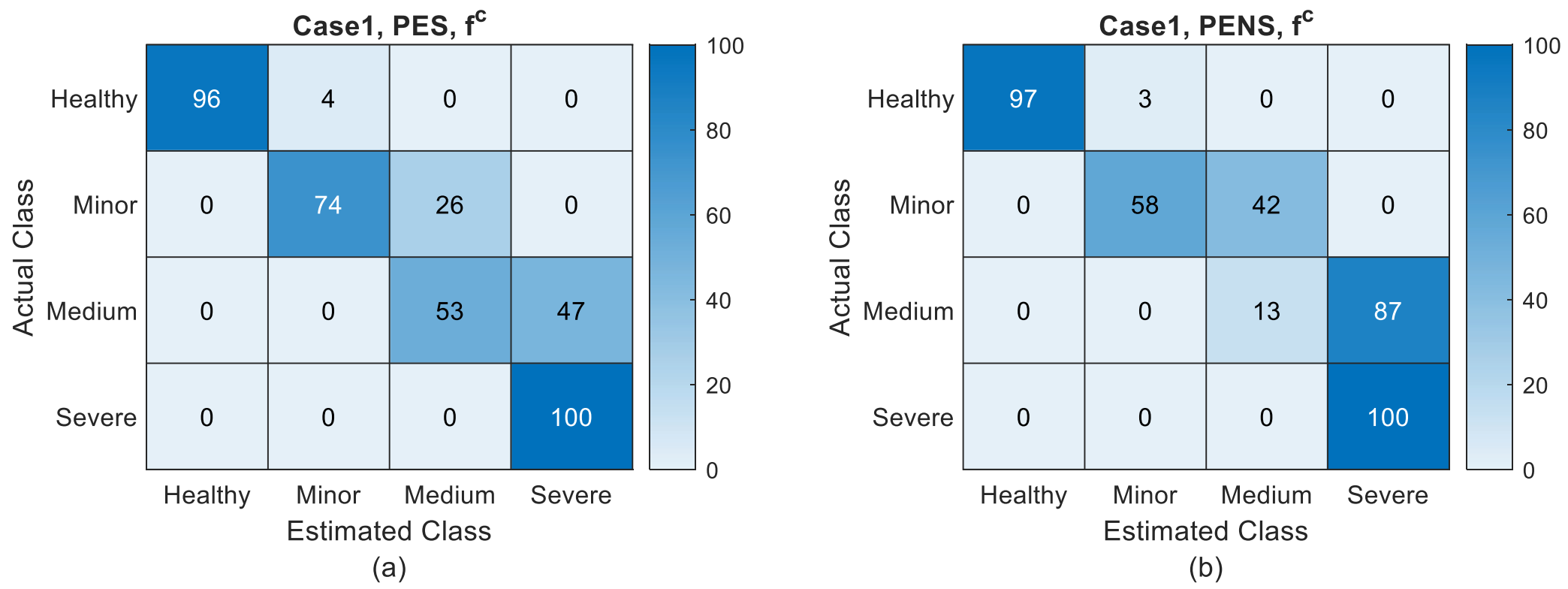

**Figure 6.** Confusion matrix for $f^c$ based on the PES (**a**)/PENS (**b**) under Case 1.

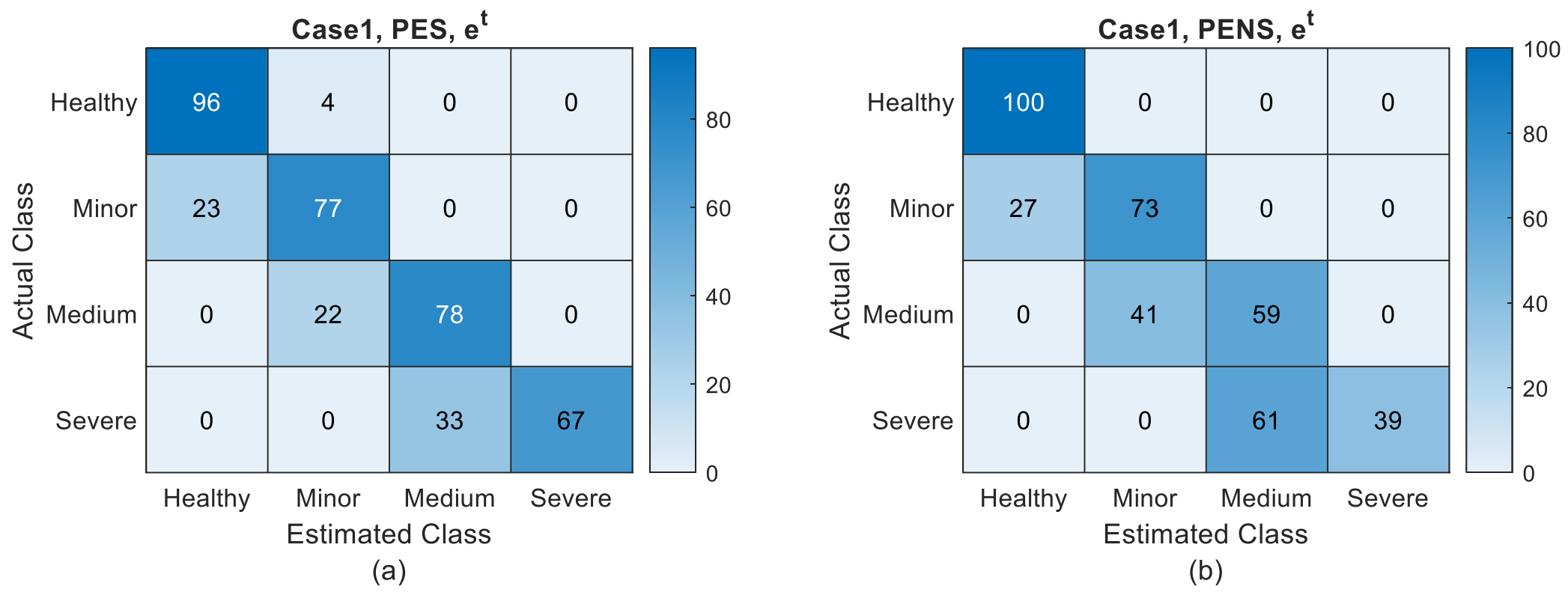

**Figure 7.** Confusion matrix for $e^t$ based on the PES (**a**)/PENS (**b**) under Case 1.

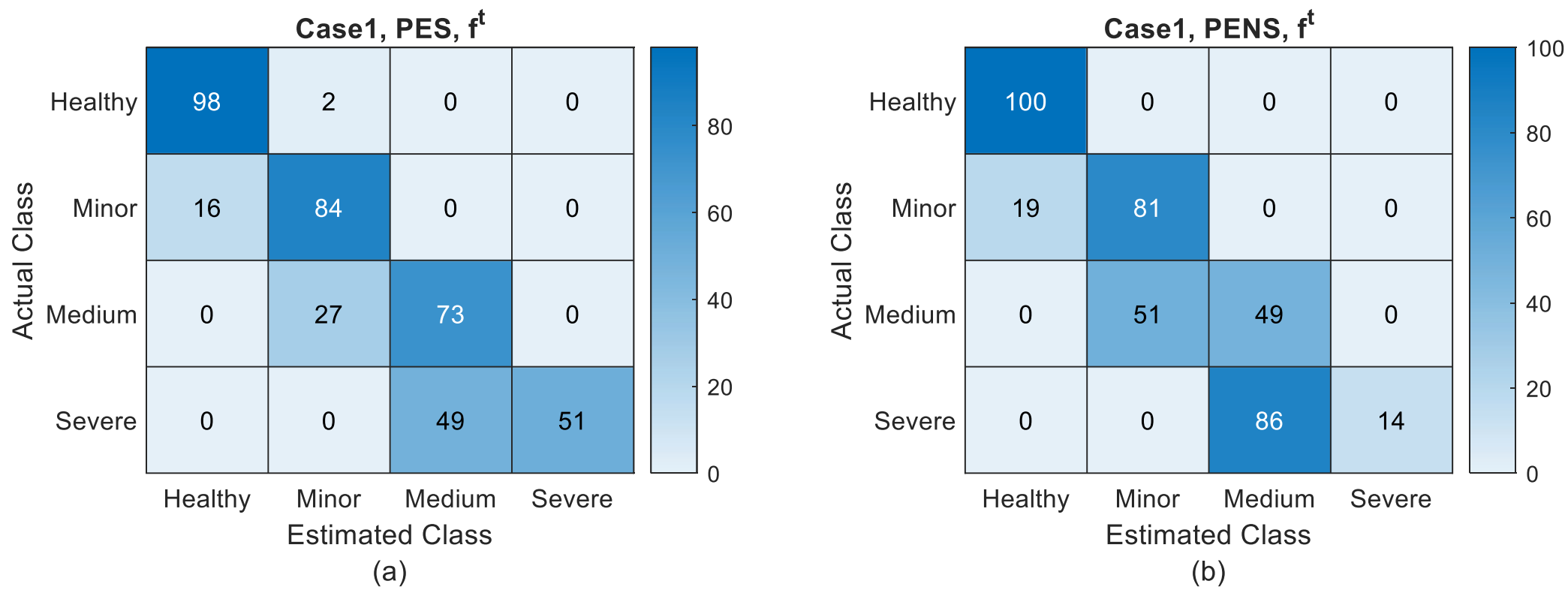

**Figure 8.** Confusion matrix for $f^t$ based on the PES (**a**)/PENS (**b**) under Case 1.

However, for minor, medium, and severe fault conditions, the FDI systems that utilize PESs exhibit markedly superior performance as compared to those using PENSs. This superiority is evidenced by higher detection rates as well as lower false positive and negative rates, particularly in the context of higher severe faults. Consequently, the simulation results from Case 1 substantiate the conclusion in Section 5 that shaft power information can improve the estimation accuracy of each state and health parameter, resulting in an enhancement in the performance of the FDI scheme.

Next, we analyze the changes in system measurements, including the state variable (rotor speed) and other measured parameters, during gas-path faults. Taking the degradation profile shown in Figure 4, we set the shaft power output to remain constant and examine the resulting system measurements as illustrated in Figure 9. The red lines represent raw measurements while the blue lines show the moving average values to better illustrate the trends, as the original measurements (red) are affected by state and measurement noise.

From the figure, we can observe that the rotor speed (state variable) exhibits minimal variation. This is because APU gas generators typically operate under fixed-speed control logic at specific operating conditions [64]. As mentioned in the model description at the beginning of this section, a PI controller was implemented to maintain the nominal speed.

Examining the system parameters in Equations (18) and (19) and the degradation curves in Figure 4, we notice that for this particular degradation profile, the health parameters' effects on $T_3$ largely cancel each other out, resulting in minimal $T_3$ variation. However, this limited variation is specific to this sample's degradation rate and should not be considered universal.

For $P_3$ and $T_5$, since their influences cannot be mutually canceled, we observe significant changes after degradation. The degraded compressor performance results in reduced compression capability, leading to lower pressure ratio for the same temperature rise. This decreased gas work capability means that to maintain the same power output, higher temperatures and increased fuel consumption are required. Consequently, we observe decreased compressor outlet pressure $P_3$ and elevated turbine outlet temperature $T_5$. For more detailed analysis of degradation effects on system performance, readers are referred to the relevant literature [30,32,38,51].

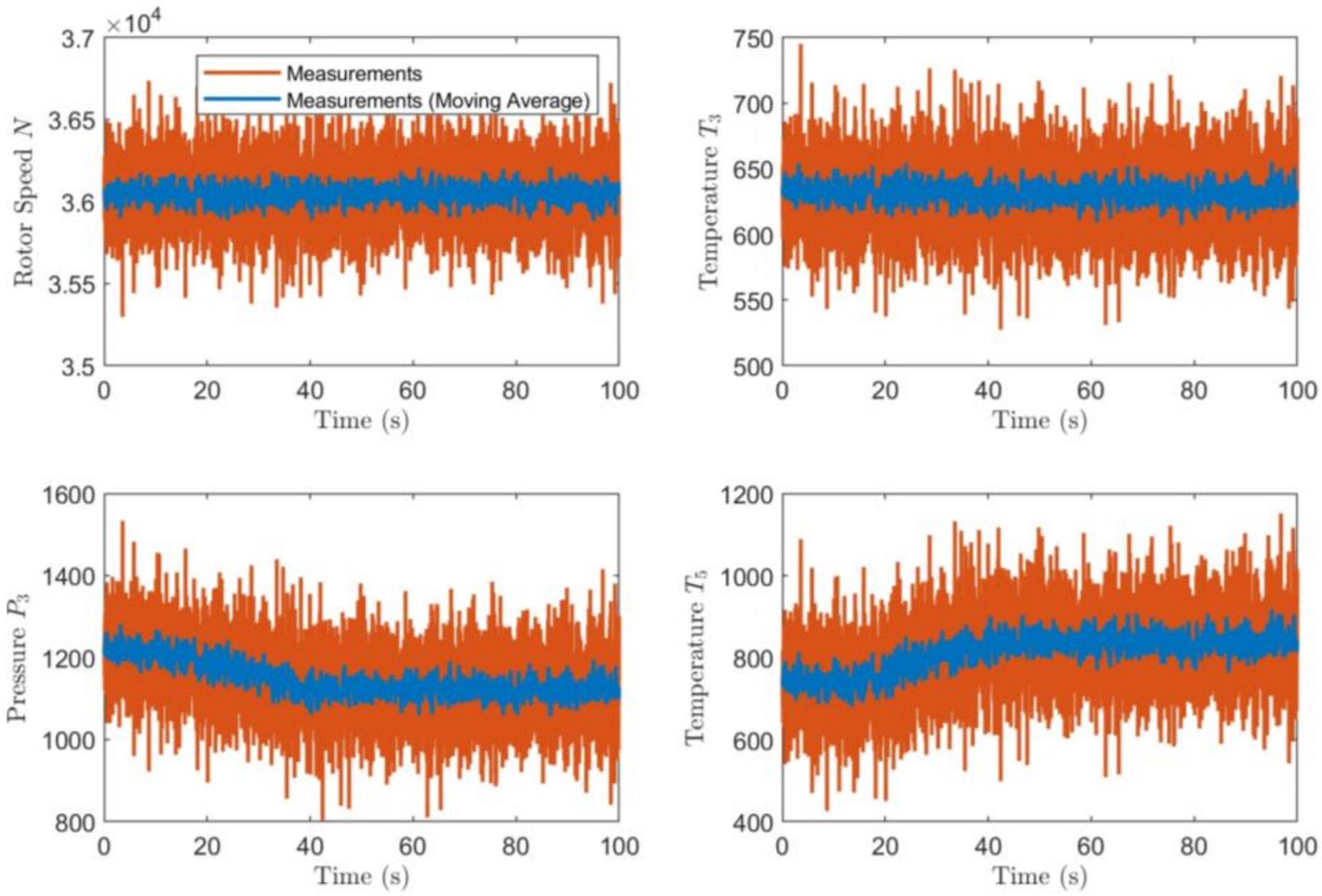

**Figure 9.** System measurements under gas-path faults.

***Case 2***: We now reduced the process noise $w^{gt}$ to 0.1% of the reference steady-state values of the system states, while keeping all other parameters the same as in Case 1. Subsequently, we also conducted 100 Monte Carlo simulation runs for each condition, as shown in Table 1. Figure 10 illustrates the dynamic estimation results from a representative simulation sample. The estimation accuracy of the PES and PENS, along with corresponding FDI results, are provided in Table 3 and Figures 11–14.

**Table 3.** Comparison of accuracy between the PES and PENS under Case 2.

| | RMSE of PES | RMSE of PENS | Accuracy Improvement of the PES over PENS |
|---|---|---|---|
| $N_L$ | 63.34 | 121.7 | 47.97% |
| $e^c$ | $1.318\times10^{-3}$ | $3.322\times10^{-3}$ | 60.33% |
| $f^c$ | $3.898\times10^{-4}$ | $1.236\times10^{-3}$ | 68.47% |
| $e^t$ | $2.326\times10^{-3}$ | $7.075\times10^{-3}$ | 67.13% |
| $f^t$ | $2.978\times10^{-3}$ | $9.328\times10^{-3}$ | 68.07% |

Figure 10 and Table 3 demonstrates the estimation accuracy and FDI capabilities in scenarios both with and without access to shaft power information, given a more accurate model. Comparing Table 3 with Table 2 shows that the accuracy of the PES increases with the use of a more accurate model, while the accuracy of the PENS remains nearly unchanged under the same conditions. This leads to a larger accuracy advantage for the PES over the PENS. It can be inferred that increased accuracy in the model state transition representation enhances the positive impact of power shaft information on the accuracy of health parameter estimation.

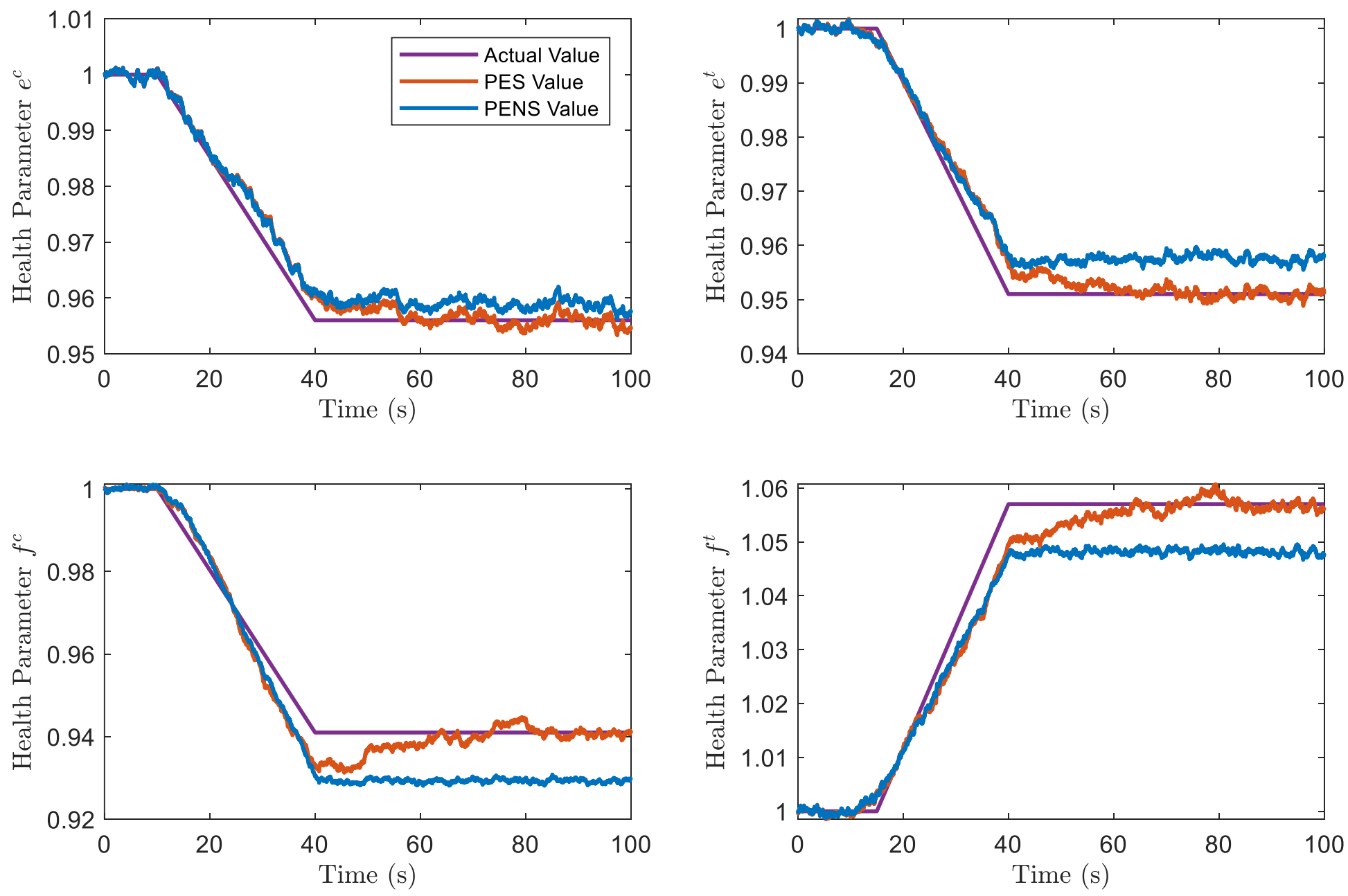

**Figure 10.** Case 2 dynamic estimation results for PES and PENS.

Note that a more accurate state transition representation leads to a more effective utilization of power shaft information, resulting in improved prior estimations and, ultimately, enhanced overall accuracy and FDI performance. However, in scenarios lacking shaft power information, the improvement in the accuracy of the state transition representation does not significantly enhance the estimation accuracy. Without the shaft power information, state transition representation plays a limited role in estimating states and health parameters, which primarily rely on the observation equation. Consequently, simulation results for Case 2 show that a more accurate model enhances the positive impact of power shaft information on the accuracy of achievable health parameter estimation results.

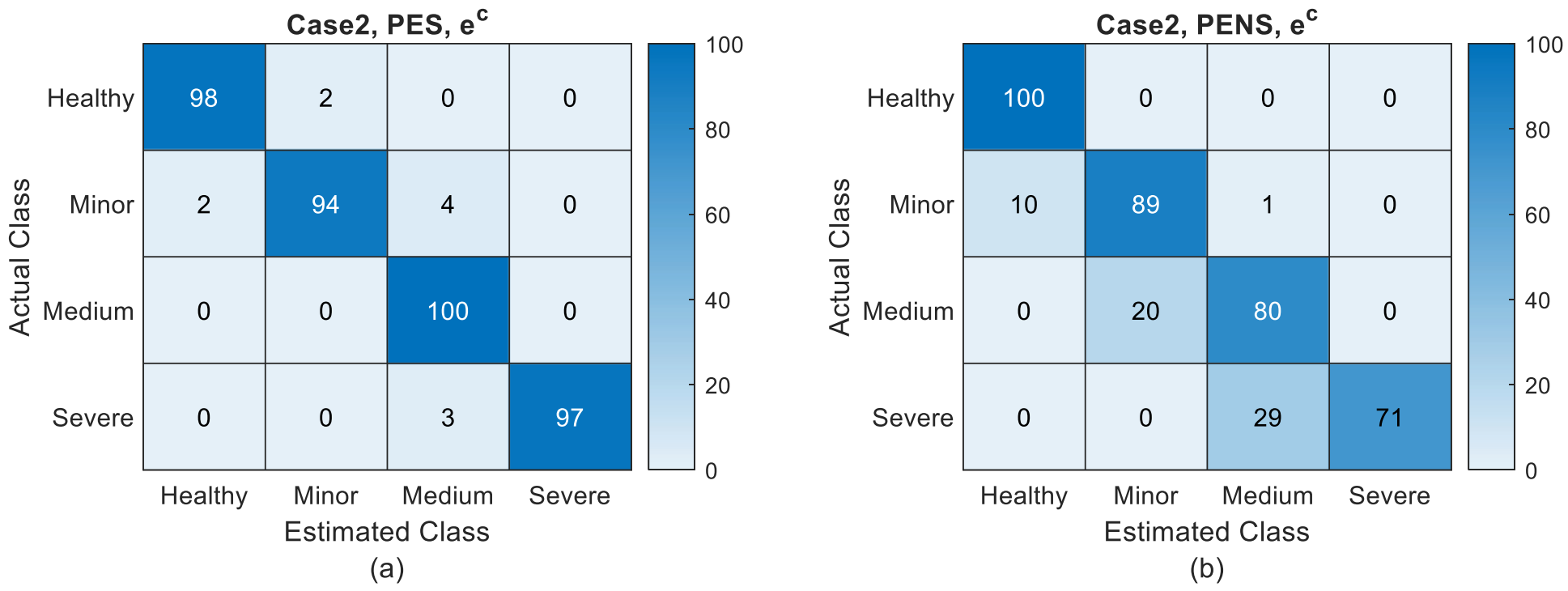

**Figure 11.** Confusion matrix for $e^c$ based on the PES (**a**)/PENS (**b**) under Case 2.

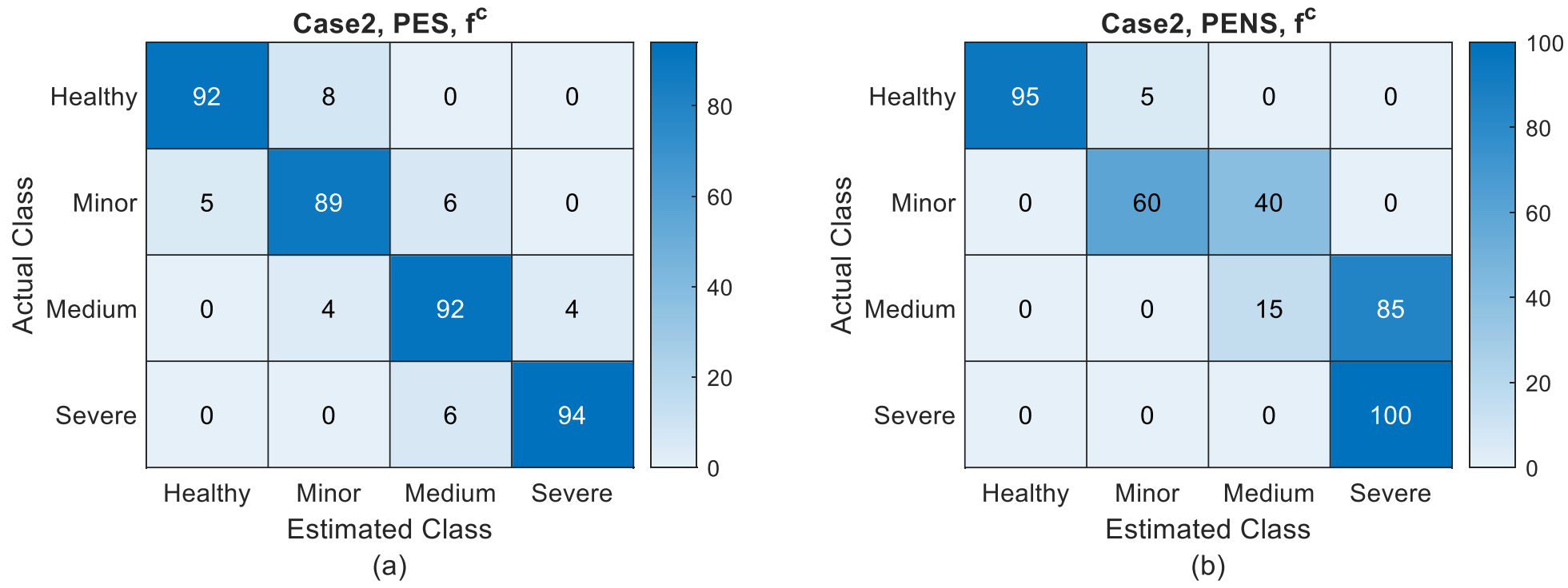

**Figure 12.** Confusion matrix for $f^c$ based on the PES (**a**)/PENS (**b**) under Case 2.

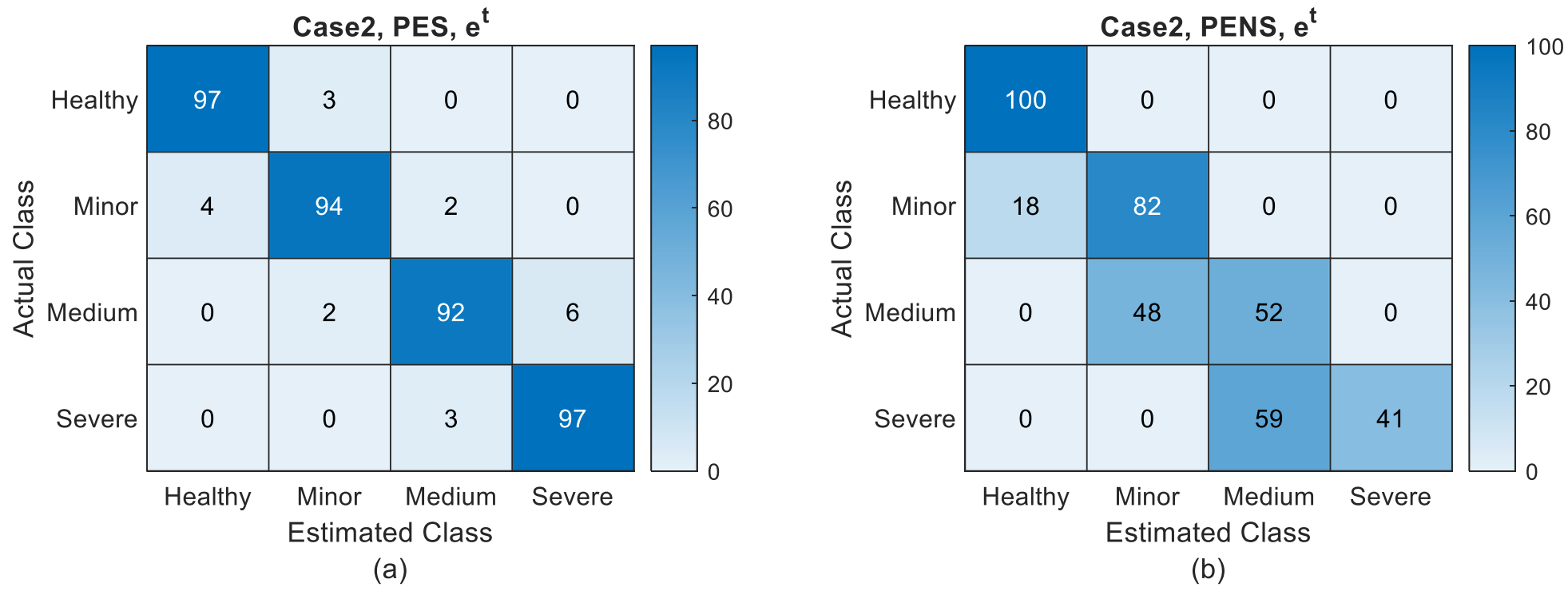

**Figure 13.** Confusion matrix for $e^t$ based on the PES (**a**)/PENS (**b**) under Case 2.

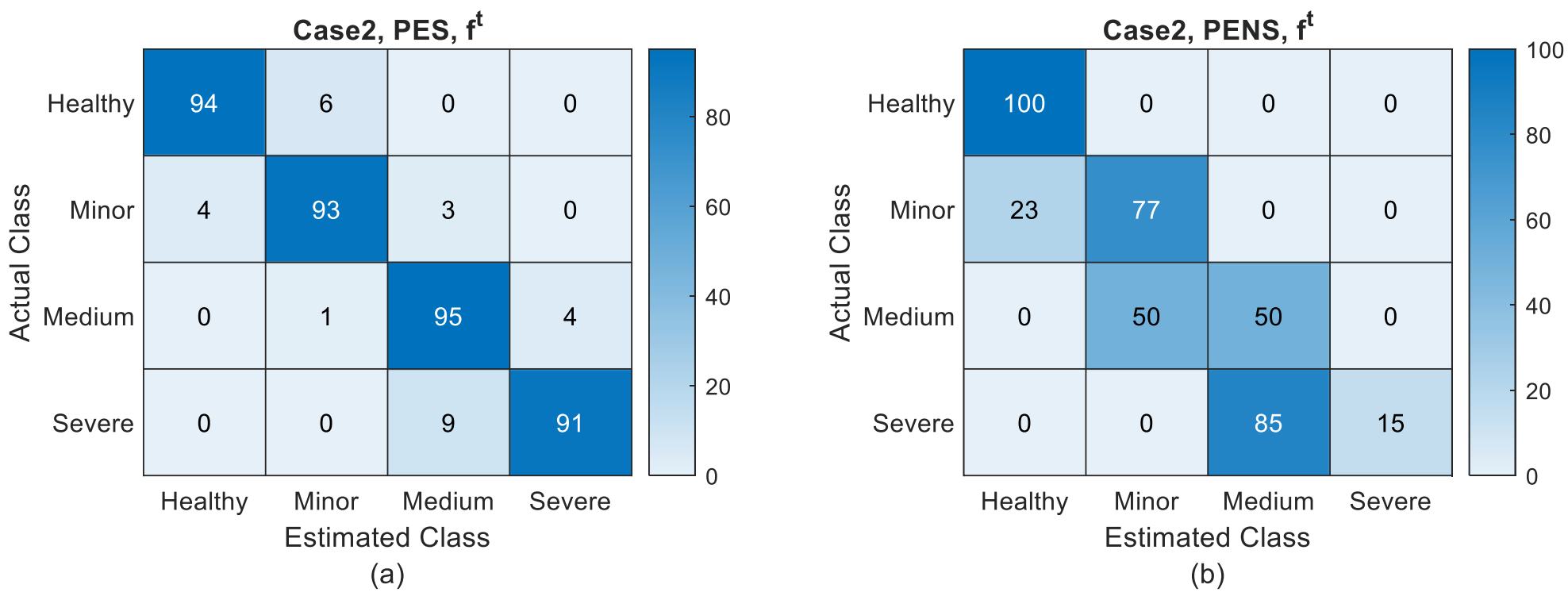

**Figure 14.** Confusion matrix for $f^t$ based on the PES (**a**)/PENS (**b**) under Case 2.

***Case 3*:** We assume that degradation of the compressor and turbine simultaneously occurs in terms of efficiency and flow, and there exists a known proportional relationship as given by Equation (22) [57]. Given that $k^{gt-c}, k^{gt-t}$ are known fixed values, the description of the gas generator gas-path degradation can be reduced from the four independent health parameters $e^{gt-c}, f^{gt-c}, e^{gt-t}, f^{gt-t}$ to the two independent variables $e^{gt-c}, e^{gt-t}$, as specifically shown in Equation (22). Next, we set the same noise level as in Case 2 and keep all other parameters unchanged. Subsequently, we also conducted 100 Monte Carlo simulation runs for each condition, as shown in Table 1. Figure 15 illustrates the dynamic estimation results from a representative simulation sample. The estimation accuracy of the PES and the PENS, along with the corresponding partial FDI results, are detailed in Table 4 and Figures 16 and 17.

$$\begin{aligned} f^{gt-c} &= 1 - k^{gt-c}(1 - e^{gt-c}) \\ f^{gt-t} &= 1 + k^{gt-t}(1 - e^{gt-t}) \end{aligned} \tag{22}$$

where $k^{gt-c}, k^{gt-t}$ respectively, denote the known fixed correlation coefficients between the efficiency factor and the flow factor of the compressor and turbine, respectively.

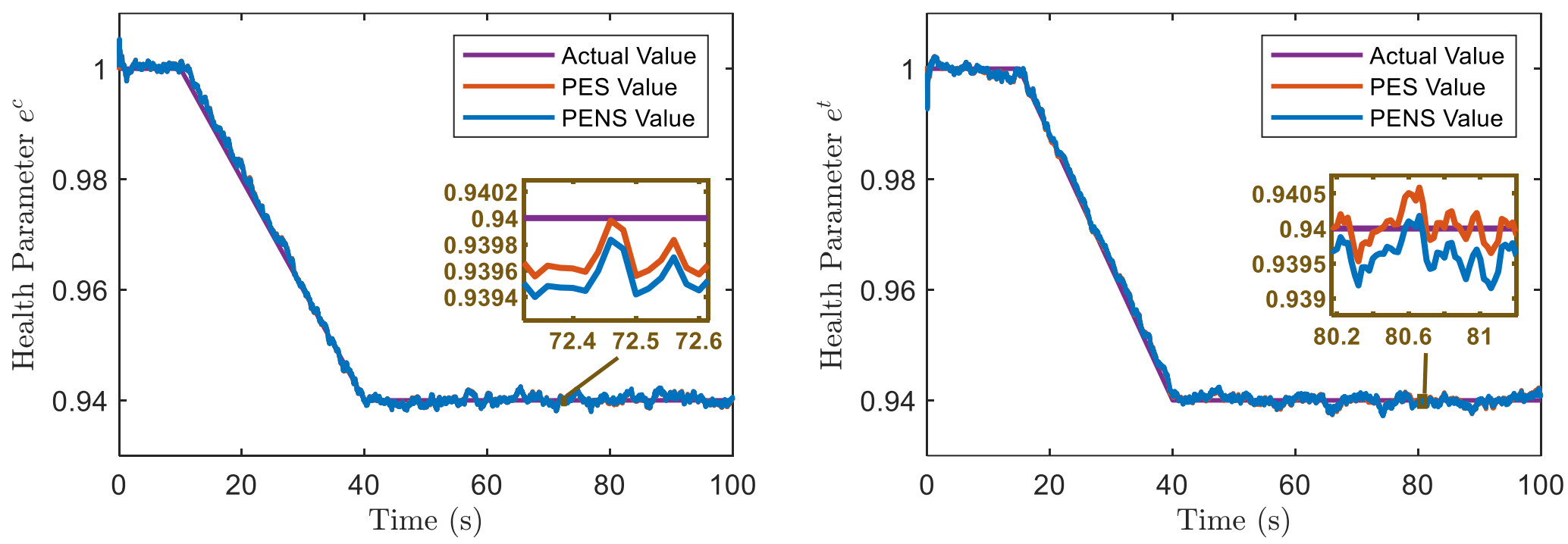

**Figure 15.** Case 3 dynamic estimation results for PES and PENS.

Figures 15–17 provide estimation accuracy and FDI performance when fewer health parameters are estimated by using the same number of measurements, indicating a relatively higher sufficiency of information for estimating health parameters. In Table 4, it follows that the estimation accuracies of the PES and PENS are nearly the same. This similarity can be attributed to the fact that estimating fewer parameters while maintaining the same number of measurements results in relatively less uncertainty in estimations. This diminished uncertainty arises from the assumption in Case 3 of a determined relationship among health parameters, thereby narrowing the possible distribution of parameters given the same measurements. Consequently, when uncertainty is lower, the contributions of shaft power estimates from the starter/generator to enhancing the accuracy of health parameter estimation are minimal. Furthermore, similar FDI performance based on both the PES and PENS, as per Figures 16 and 17, further supports our conclusion.

**Table 4.** Comparison of accuracy between the PES and PENS under Case 3.

| | RMSE of PES | RMSE of PENS | Accuracy Improvement of the PES over PENS |
|---|---|---|---|
| $N_L$ | 61.04 | 117.1 | 47.88% |
| $e^c$ | $8.053\times10^{-4}$ | $8.068\times10^{-4}$ | 0.1947% |
| $e^t$ | $8.507\times10^{-4}$ | $8.535\times10^{-3}$ | 0.3351% |

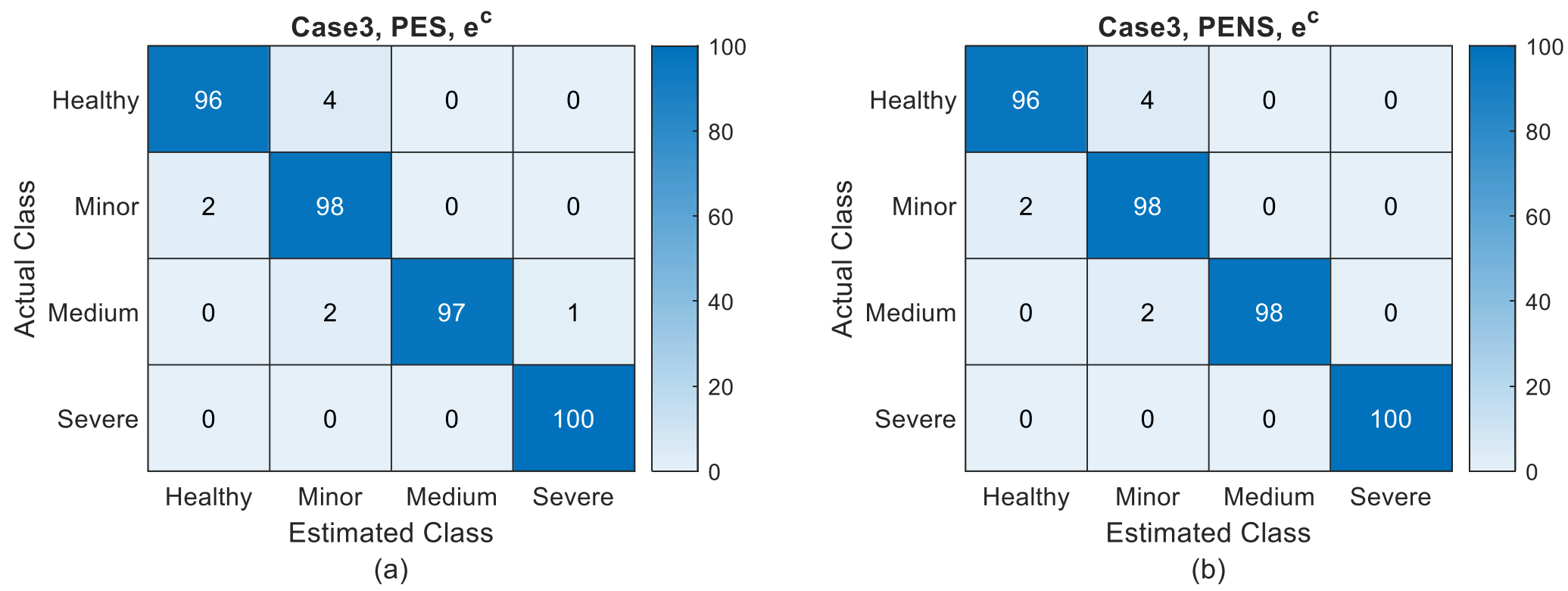

**Figure 16.** Confusion matrix for $e^c$ based on the PES (**a**)/PENS (**b**) under Case 3.

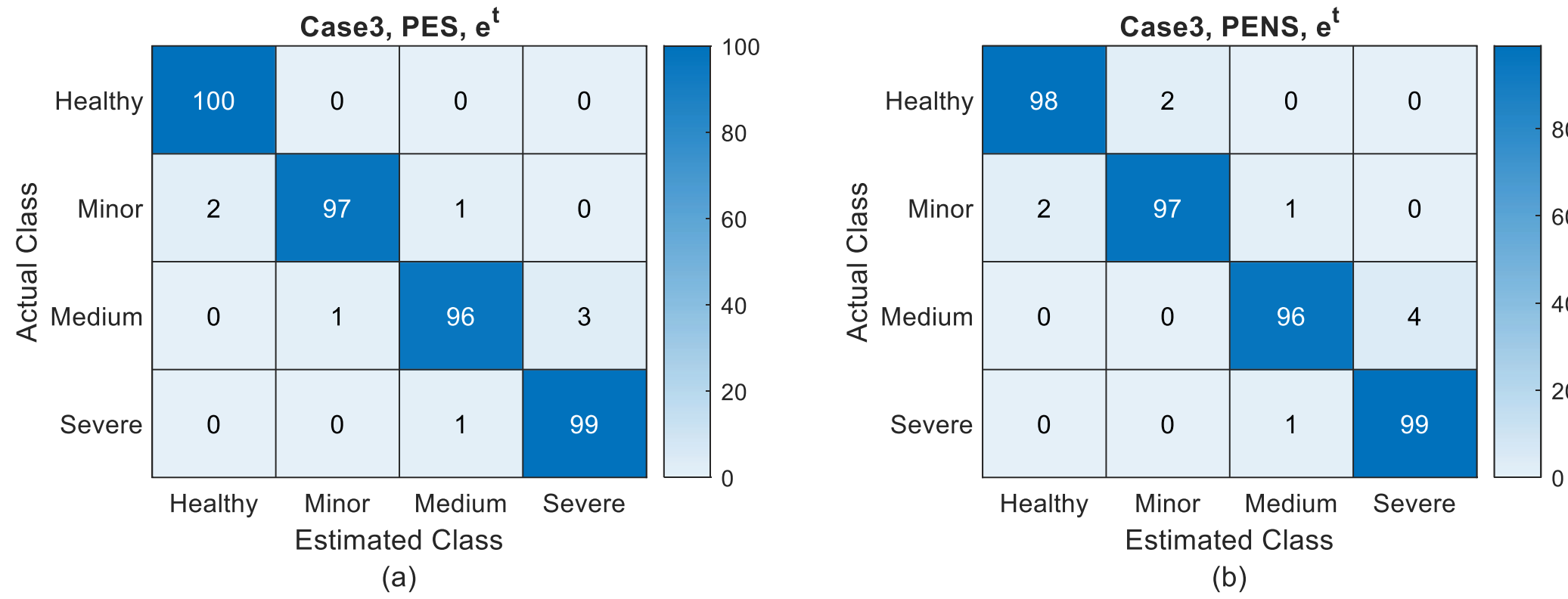

**Figure 17.** Confusion matrix for $e^t$ based on the PES (**a**)/PENS (**b**) under Case 3.

Consequently, comparisons between Cases 3 and 2 indicate that the contribution of shaft power information in improving the estimation accuracy of system states and health parameters becomes more significant when more health parameters are estimated.

In Table 5 below, we present the macro averages of precision, recall, and F1 score for FDI based on the PES and PENS across different health conditions in Case 1, Case 2, and Case 3. From the data in Table 5 for Case 1, it is evident that the PES outperforms PENS due to its use of precise shaft power information. Furthermore, the data in Table 5 for Case 2 show that as the model accuracy improves, the advantages of the PES become more pronounced. However, data in Table 5 for Case 3 also indicate that when the number of health parameters to be estimated decreases, the importance of additional shaft power information diminishes due to reduced system uncertainty, resulting in similar performance between the PES and PENS.

**Table 5.** Performance metrics for FDI based on PES and PENS.

| Simulation Cases | Estimates Used | Precision | Recall | F1 Score |
|---|---|---|---|---|
| Monte Carlo Case 1 | PES | 0.8344 | 0.8193 | 0.8162 |
| | PENS | 0.7541 | 0.7131 | 0.6927 |
| Monte Carlo Case 2 | PES | 0.9445 | 0.9437 | 0.9438 |
| | PENS | 0.7453 | 0.7044 | 0.6850 |
| Monte Carlo Case 3 | PES | 0.9790 | 0.9788 | 0.9788 |
| | PENS | 0.9778 | 0.9775 | 0.9775 |

It follows that the relative advantage of the PES over PENS lies in its significant accuracy benefits when measurement information is relatively insufficient, the number of health parameters to be estimated is large, or the model accuracy is high. On the other hand, the relative advantage of the PENS is that it does not require additional precise shaft power information provided by a generator, making it easier to deploy. It also performs well when model accuracy is not high or the number of health parameters to be estimated is small. Therefore, the choice between the PES and PENS should be based on the specific scenario and application under consideration.

The above Monte Carlo simulation runs and related analyses have confirmed our conclusion that incorporating shaft power information can enhance the estimation accuracy of each state and health parameter, thereby improving the FDI performance. These improvements are especially pronounced with a more accurate model or when more health parameters are involved, as discussed in Section 5.

In this study, linear degradation patterns are employed for validation purposes. This choice is justified as our focus is on fault estimation rather than fault prediction, and the proposed estimation method is not significantly dependent on the specific form of the degradation curve. As long as the degradation process is gradual, the estimation accuracy remains robust across different degradation patterns. This is evidenced through our Monte Carlo simulations, where various degradation rates are tested despite the linear form. While more complex stochastic degradation models exist in the reliability literature, linear degradation models are commonly used in fault estimation studies [30,65,66] due to their ability to effectively validate estimation algorithms.

It should be noted that the above model used for simulation validation is based on a single-shaft all-electric APU gas generator gas-path fault model referenced from the currently prevalent Hamilton Sundstrand APS 5000. However, our analysis and conclusions are derived through mathematical analysis and are of general applicability. The analytical framework and conclusions presented herein are valid for any aviation gas turbine and associated faults that conform to the model specifications and assumptions established in Section 3, including all-electric APUs with alternative configurations and hybrid electric aircraft power systems that exhibit analogous structural characteristics.

Furthermore, it should be emphasized that this study investigates the theoretical upper bounds of estimation capability rather than specific implementation approaches. While our theoretical derivation and simulation validation employ Kalman filtering, we utilize it solely for its optimal estimation properties to establish the best achievable estimation accuracy under ideal conditions, rather than as a proposed implementation method. Our analysis demonstrates that when a system satisfies our model assumptions, the theoretical upper bound of estimation accuracy with shaft power information is higher than that without such information, and this advantage becomes more pronounced when the model is more accurate or when more health parameters need to be estimated. This fundamental conclusion about theoretical limits is independent of practical implementation aspects such as specific estimation methods, system identification, or parameter acquisition procedures.

## 7. Conclusions

This study presents a comprehensive analysis of the opportunities and challenges associated with the FDI in the context of APU electrification. We have proposed an innovative methodology to enhance both state and fault estimation of the gas generator in an all-electric APU by leveraging generator information. Our rigorous analysis demonstrates that the incorporation of starter/generator shaft power information significantly improves the optimal estimation accuracy for both the states and the health parameters of the gas generator. This enhancement becomes particularly evident when employing more accurate models or when estimating an expanded set of health parameters. The effectiveness of our proposed methodology has been thoroughly validated through extensive simulation studies.

Our research findings have provided a perspective as to why joint fault diagnosis of the starter/generator and gas generator is necessary for all-electric APUs, how to conduct such joint diagnosis, what is its effectiveness, and what factors affect its effectiveness. These offer guidance and lay the theoretical foundation for the development of current fault diagnosis algorithms for gas generators in all-electric APUs. In addition to their application in all-electric APUs, our research findings can also be directly applied to future hybrid electric aircraft, including turboelectric and series hybrid electric architectures. These systems, which are based on aviation gas turbines dedicated to electrical power generation, share nearly identical principles and structures with all-electric APUs. Moreover, our theoretical framework has the potential to be extended to other electro-mechanical systems with similar structures, such as REEVs and wind turbines. In our future work, we will further investigate new challenges that are posed by all electrification of APUs as discussed in this paper. We aim to explore joint fault diagnosis algorithms for starters–generators and gas generators that are capable of simultaneously exploiting new opportunities and addressing new challenges.

**Author Contributions:** Conceptualization, H.M.; methodology, H.M.; software, H.M.; validation, H.M.; formal analysis, H.M.; investigation, H.M.; resources, K.K. and Y.G.; data curation, H.M.; writing—original draft preparation, H.M.; writing—review and editing, K.K. and Y.G.; visualization, H.M.; supervision, K.K. and Y.G.; project administration, K.K. and Y.G.; funding acquisition, K.K. and Y.G. All authors have read and agreed to the published version of the manuscript.

**Funding:** This research was funded by China Scholarship Council under Grant 201906290241.

**Data Availability Statement:** The original contributions presented in this study are included in the article. Further inquiries can be directed to the corresponding author.

**Conflicts of Interest:** The authors declare no conflicts of interest.

## Abbreviations

The following abbreviations are used in this manuscript:

| | |
|---|---|
| AEA | All-electric aircraft |
| MEA | More-electric aircraft |
| APU | Auxiliary power unit |
| CBM | Condition-based maintenance |
| LSTM | Long short-term memory |
| SVR | Support vector regression |
| FDI | Fault detection and isolation |
| EKF | Kalman filter |
| SAE | Stacked autoencoder |
| SVDD | Support vector data description |
| BWFSG | Brushless wound-field synchronous generator |
| REEV | Range-extended electric vehicle |
| PENS | Posterior estimate with no shaft power information |
| PES | Posterior estimate with shaft power information |
| MPES | Modified posterior estimate with shaft power information |
| CEE | Covariance of estimation error |
| RMSE | Root mean square error |